\documentclass[article, aps,preprint,preprintnumbers,amsmath,amssymb]{revtex4-1}
\usepackage{epsfig}% Include figure files
\usepackage{dcolumn}% Align table columns on decimal point
\usepackage{graphicx}
\usepackage{bm}
\usepackage{hyperref}
\usepackage{color}  % colored text to highlight revisions 
\usepackage{setspace}
\usepackage{appendix}
\RequirePackage{lineno}

\newcommand{\wrt}{with respect to\hspace{0.1cm}}

\newcommand{\emf}{electromagnetic\hspace{0.1cm}}

\definecolor{Gray}{gray}{0.85}
\definecolor{LightCyan}{rgb}{0.88,1,1}
\begin{document}

\title{{\color{black}Finite $\beta$} Effects on Short Wavelength Ion Temperature Gradient Modes}

\author{M. Jagannath\textsuperscript{1}}
\email{jaga.physics94@gmail.com; jagannath.mahapatra@ipr.res.in}
\author{ J. Chowdhury\textsuperscript{2}}
\author{ R. Ganesh\textsuperscript{1}}
%\email{ganesh@ipr.res.in} 
\author{L. Villard\textsuperscript{3}}

\affiliation{\textsuperscript{1}Institute for Plasma Research, Bhat, Gandhinagar, 382428, India}
\affiliation{\textsuperscript{2}University of Warwick, Coventry CV4 7AL, UK}
\affiliation{\textsuperscript{3}Swiss Plasma Center, EPFL, 1015 Lausanne, Switzerland}
%
%\author{M. Jagannath}%\textsuperscript{1}}
%\email[ E-mail me at : ]{jaga.physics94@gmail.com/jagannath.mahapatra@ipr.res.in}
%\affiliation{Institute for Plasma Research, Bhat, Gandhinagar, 382428, India}
%
%\author{ J. Chowdhury}%\textsuperscript{2}}
%\affiliation{University of Warwick, Coventry CV4 7AL, UK}
%
%\author{ R. Ganesh}%\textsuperscript{1}} 
%\affiliation{Institute for Plasma Research, Bhat, Gandhinagar, 382428, India}
%
%\author{L. Villard}%\textsuperscript{3}}
%\affiliation{Swiss Plasma Center, EPFL, 1015 Lausanne, Switzerland}
\begin{abstract}
The electromagnetic effect is studied on the short wavelength branch of the ion temperature gradient mode in the linear regime for the first time using a global gyrokinetic model. The short wavelength ion temperature gradient mode growth rate is found to be reduced in the presence of electromagnetic perturbations {\color{black}at finite plasma $\beta$}. The effect on real frequency is found to be weak. The threshold value of $\eta_i$ is found to increase for the mode as the magnitude of $\beta$ is increased. \textcolor{black}{The global mode structure of the short wavelength branch
of the ion temperature gradient mode is compared
with the conventional branch. {\color{black}The magnetic character of the mode, measured as} the ratio of mode average square values of
electromagnetic potential to electrostatic potential, is found to increase
with increasing values of the plasma $\beta$. The mixing length estimate
for flux shows that the maximum contribution still comes from the long
wavelengths modes. The magnitude of \textcolor{black}{the} flux decreases with increasing $\beta$.}

\end{abstract}
\maketitle

\section{Introduction}
\indent
It is now well known that magnetically confined plasma{\color{black}s} bear the prospects of a future fusion reactor.
The progress in this endeavor, however, is hamstrung by technological and physics issues. 
With the available technology, the success of such a goal depends on understanding 
the nature of the instabilities and controlling them. The Larmor radius scale instabilities, which are also 
\textcolor{black}{called} microinstabilities, cause anomalous transport leading to  expulsion of heat and particles
from the system. This is undesirable and has to be avoided or at least reduced. 
These microinstabilities feed on the free energy available {\color{black}in the inhomogeneities} in 
temperature and density of the particles. The ion temperature
gradient (ITG)~\cite{itg1, itg2, itg3, itg4, itg5} mode, trapped electron mode (TEM)~\cite{tem1, tem2, tem3, 
tem4, tem5, tem6}, and universal drift instabilities~\cite{unv1, unv2, unv3}, are some of the examples of such 
unstable modes {\color{black}at} the ion scale while the electron temperature gradient mode (ETG)~\cite{etg1, etg2, 
etg3, etg4} is another class of instabilities {\color{black}at} the electron scale. {\color{black}Finite} $\beta$ plasma{\color{black}s} also give rise to electromagnetic instabilities such as microtearing modes 
(MTM)~\cite{mtm1, mtm2, mtm3, mtm4, mtm5, mtm6, mtm7, mtm8, mtm8a}
and kinetic ballooning modes (KBM)~\cite{kbm1, kbm2, kbm3, kbm4}.
Intermediate to these scales, there exists a class of instabilities driven by \textcolor{black}{a} strong 
ion temperature gradient. This mode occurs on the shorter scale than the conventional 
ITG mode mentioned above and therefore, is \textcolor{black}{called} the short wavelength ion temperature gradient ({\color{black}SWITG})
mode~\cite{sw1, sw2, sw3, sw4, sw5, sw6, sw7, sw8, sw9}. Usually\textcolor{black}{,} the ion temperature gradient 
driven mode exists for \textcolor{black}{$k_{\perp}\rho_{Li}\le 1$}. However, when the gradient scale lengths are very short,
one can observe another branch of ITG mode which becomes unstable at scales \textcolor{black}{$k_{\perp}\rho_{Li}\textgreater 1$}.\\
\indent 
With the progress in the tokamak fusion research, tokamaks are now able to operate in advanced scenarios.
Some of such scenarios offer very strong gradients. For example in the region of internal transport 
barrier{\color{black}s}, the gradients can be very strong~\cite{tb1}. In H mode plasma as well, the profiles turn
very steep, resulting in very strong gradients~\cite{hm}. 
Apart from other modes, the shorter wavelength mode might be very unstable 
with the $k$ spectrum extending well beyond the conventional modes~\cite{mrl1} 
for parameters relevant to tokamak experiments. \\
\indent
The current and future tokamak machines confine plasma at a higher pressure which is expressed 
in $\beta$. Therefore, the modification of the mode properties in the presence of finite 
$\beta$ has to be taken into consideration. The $\beta$ effect turns out to be 
very important for both inherently electrostatic and electromagnetic modes. While the electromagnetic 
instabilities arise and become stronger with increasing $\beta$, the same finite-$\beta$ effect  {\color{black}suppresses}  
the electrostatic instabilities. In the presence of fast ion, high $\beta$ 
leads to the reduction in the profile stiffness observed in many experiments and 
confirmed in subsequent gyrokinetic studies~\cite{citrin}.
The role of electromagnetic perturbations on drift modes has been studied extensively. 
While \textcolor{black}{the} electromagnetic perturbation \textcolor{black}{is} observed to give rise to instabilities such as \textcolor{black}{the} kinetic ballooning mode (KBM)~\cite{kbm1, kbm2, kbm3, kbm4}, tearing and microtearing modes~\cite{mtm1, mtm2, mtm3, 
mtm4, mtm5, mtm6, mtm7, mtm8}, etc., \textcolor{black}{the} same is found to stabilize some other drift modes such 
as \textcolor{black}{the} ITG mode, trapped electron mode, 
universal drift modes, etc~\cite{itgem1, itgem2, itgem3, itgem4, unv2}. The \textcolor{black}{stabilizing} 
effect of the electromagnetic perturbation on the ITG mode can be attributed to the field line bending induced by the \emf perturbations. Although \emf effects on ITG mode are well known, the effect of \textcolor{black}{the} \emf perturbation on the SWITG mode is 
not known so far \textcolor{black}{in detail~\cite{sw6}}. \textcolor{black}{Previously known finite $\beta$ studies on the SWITG mode are in slab geometry~\cite{sw5} and in the local limit~\cite{sw6}}. We know that {\color{black}the} SWITG mode is inherently an ion mode~\cite{sw6}, and the magnetic shear has a stabilizing effect~\cite{sw5}. The dependence of the mode on $T_e/T_i$, toroidicity, magnetic shear, safety factor, 
etc., are studied in Refs~\cite{sw7, sw8}. The role of $E\times B$ shear has been studied in Ref.~\cite{sw5}.
Thus the properties of the SWITG mode have been extensively studied in the electrostatic limit. However, in the presence of $\beta$ or electromagnetic perturbation, which corresponds to a 
more realistic scenario of tokamak experiments, the \textcolor{black}{detail} properties of the SWITG mode have not been investigated so far \textcolor{black}{using a global model}. To \textcolor{black}{close} this gap in the knowledge of SWITG, {\color{black}in the present 
work}, we explore the properties of the SWITG mode in the presence of finite $\beta$. 
%Also note that, for high $n$ modes the local calculations might be appropriate; but for low and intermediate $n$ modes, the mode can span over several mode rational surfaces
%with mode width multiple times the Larmor radius. The global profile effects become
%important in such cases. {\color{black}It is clear from previous simulations ~\cite{sw8} that although the SWITG modes
%appears at high $n$, the mode is quite global and local calculations overestimate the
% growth rates
%and real frequencies in certain cases}. This underscores the importance of the use of {\color{black}a} global code in the study of the SWITG mode.
 For this \textcolor{black}{reason},
we  use the electromagnetic version EM-GLOGYSTO~\cite{gloria3, kbm4, rg_prl} of the  global 
gyrokinetic linear model GLOGYSTO~\cite{sb_cyl, sb_tor, sb_thesis}. This code has been extensively used 
to study electrostatic and electromagnetic modes. For example, in the electrostatic limit the 
code is used to study ITG~\cite{sb_tor, itg4}, TEM~\cite{tem5}, SWITG~\cite{sw8} modes while in the electromagnetic limit 
KBM~\cite{gloria3, kbm4, rg_prl}, and MTM~\cite{mtm6, mtm6a} modes 
are studied in detail.  
The present manuscript is arranged as follows. Section II describes the model briefly. In Section III, 
the profiles and parameters used in the present simulations, results observed in the presence of {\color{black}finite} $\beta$ are described. \textcolor{black}{Finally}, the results 
are summarized in Section IV. 
\vspace{-0.2in}
\section{The Simulation Model}
%\noindent
The EM-GLOGYSTO code is a global spectral code that calculates the real frequency and growth rates
of unstable modes for a given equilibrium using the Nyquist method and also gives the {\color{black}eigen}mode 
structure. Both ion and electron species are considered fully gyrokinetic. 
The equilibrium considered is circular and axisymmetric and can include 
the Shafranov shift. The code can treat 
trapped and passing particles separately and calculates the finite Larmor radius (FLR) effect
 to all orders. It includes all kinetic effects including Landau damping physics. The model includes 
both transverse and compressional perturbation in the electromagnetic limit. In the present 
study, however, we consider $\phi$ and $A_{\parallel}$ components only.
\noindent
Before proceeding to present the results we briefly summarize the simulation model in this
section. A greater detail of the model can be found in Refs.~\cite{sb_cyl, sb_tor}. 
For electromagnetic calculations, readers are referred to Refs.~\cite{gloria3, kbm4}.
The perturbed density for a species $j$ can be expressed as the sum of adiabatic and nonadiabatic parts 
as follows.
\begin{eqnarray}
 {\tilde n}_{j}({\bf r};\omega) = - \left(\frac{q_{j}N}{T_{j}}\right)
\Bigg[ {\tilde\varphi} + \int d{\bf k} \exp\left({i {\bf k}\cdot{\bf r}}\right)\;
\int d{\bf v} \frac{f_{Mj}}{N}\left(\omega-\omega^{*}_{j}\right)\left(i
{\cal U}_{j}\right)
\tilde \varphi({\bf k};\omega)J^{2}_{0}(x_{Lj})\label{eq_n}
\Bigg]
\end{eqnarray}
In the above equation $q_{j}$ and $T_{j}$ are the charge and temperature for the species $j$, N stands
for the equilibrium density, \textcolor{black}{$\tilde{\varphi}$ is the perturbed electrostatic potential, ${\bf k} = {\kappa}\;{\vec{e}}_{r}+k_{\theta}\;{\vec{e}}_{\theta}+k_{\phi}\;{\vec{e}}_{\phi}$ 
where radial wave vector ${\kappa} = (2\pi/{\Delta r})k_{r}$, with $\Delta r=r_{u}-r_{l}$ defines 
the radial domain, poloidal wave vector $k_{\theta} = m/r$, toroidal wave vector $k_\phi=n/R$; and ($ k_{r},m,n$) are radial, poloidal 
and toroidal mode numbers respectively. This model uses a toroidal coordinate system with $(r,\theta,\phi)$ representing radial, poloidal, and toroidal coordinates respectively. Similarly, $(\kappa,k_\theta,k_\phi)$ represent their respective conjugate coordinates in Fourier space.} The diamagnetic drift frequency is given by $\omega^{*}_{j}=\omega_{nj}\left[1+\frac{\eta_{j}}{2}\left(\frac{v^2}{v^{2}_{thj}}-3\right)\right]$ where $\omega_{nj}=(T_{j}\nabla_{n}\ln N k_{\theta})/(q_{j} B)$, $\nabla_{n}=-rB_{p}\frac{\partial}{\partial \psi}$, and $\eta_{j}=(d\ln T_{j})/(d\ln N)$, $v_{thj}$ is the thermal velocity of species $j$. The Bessel function $J_{0}(x_{Lj})$ 
with $x_{Lj}=k_{\perp}\rho_{Lj}$, incorporates the full finite Larmor radius effect. 
Note that here $m$ and $n$ are poloidal and toroidal wave numbers, $q(s)$ is the safety factor, 
$k_{\theta}$ is {\color{black}the} poloidal wave vector, $B_{p}$ is {\color{black}the} poloidal magnetic field. $f_{Mj}$ is a local Maxwellian for species {\color{black}$j$} of mass $m_{j}$ and is given by
$$f_{Mj}(\varepsilon_{\color{black}j},\psi)=\frac{N(\psi)}{{\left(\frac{2\pi T_{j}(\psi)}{m_{j}}\right)}^{3/2}}\;
\exp \left(-\frac{\frac{1}{2}m_{{\color{black}j}}v^{2}}{T_{j}(\psi)} \right)$$
The term ${\cal U}_{j}$ represents the guiding center propagator for the passing particles where, 
\begin{eqnarray}
{i }{\cal U}_{j} = \sum_{p,p'} \frac{ J_{p}(x^{\sigma}_{tj}) J_{p'}(x^{\sigma}_{tj}) }
{\omega - \sigma k_{||}|v_{||}| - p \omega_{t} } \exp(i(p-p')(\theta-\bar\theta_{\sigma})),
\label{eq_propagator_Jp}
\end{eqnarray}
with \textcolor{black}{$x^{\sigma}_{tj}=k_{\perp}\xi^\sigma_j$, 
$\xi^\sigma_j = v_{dj}/\omega_{t}$}, 
$v_{dj} = \left(v_{\perp}^{2}/2 + v_{||}^2\right)/(\omega_{cj}R), 
\mu=\frac{v_{\perp}^2}{2B}, 
\omega_{t} = \sigma |v_{||}|/(q(s)R), 
\sigma =\pm 1$ (sign of $v_{||}$), {\color{black}$\varepsilon_j=\frac{1}{2}m_j v^{2}$}. The perpendicular and parallel wave-vectors are 
given as $k_{\perp} = \sqrt{\kappa^{2} + k^2_{\theta}}$, $k_{||} = \left[nq(s) - m \right]/(q(s)R)$.
Also note that $\bar\theta$ is given by $\tan\bar\theta = -\kappa/k_{\theta}$.
% with $s=\rho/a$, $a$ and $R$ being the minor and major radius of a tokamak.
The Bessel functions contain the effect of $\nabla$B and curvature drifts through 
arguments ($x^{\sigma}_{tj}=k_{\perp}v_{dj}/\omega_{t}$). Thus the 
 Bessel functions in Eq.(\ref{eq_propagator_Jp}) take in to effect the coupling of the flux surfaces
 and also the coupling of the  neighboring poloidal modes. 
Note that the argument of the Bessel functions $J_{p}$ in Eq.(\ref{eq_propagator_Jp}), \textcolor{black}{$x_{tj}^{\sigma}=k_{\perp}\xi^{\sigma}_j$}
also takes into account the effect of 
transit frequency $\omega_{t}$. 
% and transit harmonic orders $p$ are to be chosen accordingly. 
%\noindent
%In this form, ${\cal U}_{j}$ contains effects such as transit
%harmonic resonances, parallel velocity resonances and poloidal mode coupling. 
The quasi-neutrality condition then gives
\begin{eqnarray}\label{eq_closure}
\sum_{j} {\tilde n}_{j}({\bf r};{\omega})=0,\;\;\;\;\; 
\end{eqnarray}
In the case of electrostatic fluctuations only, this leads to an eigenvalue problem, with 
$\omega$ and  ${\tilde \varphi}$ being eigenvalues and eigenvectors and can be solved in Fourier space. 
%For a given toroidal mode number $n$ this can then be conveniently solved in the Fourier space ($\kappa,m$)
%by Fourier decomposing the potential $\tilde \varphi$ in Eq.(1) first and then
%taking the Fourier transform of $\tilde n_{j}$, to eventually obtain  a convolution matrix in Fourier space.
For fully gyrokinetic ions and electrons with only passing particles we have:
\begin{eqnarray}
\displaystyle
\sum_{{\bf k'}} \sum_{j={\rm i,e}}{\cal \hat{M}}^{j}_{{\bf k},{\bf k'}} \;\; {\tilde \varphi}_{\bf k'}
=0.
\end{eqnarray}
For an axisymmetric system, one can fix the toroidal mode number $n$, and thus ${\bf k} = (\kappa,m)$ for the
    wave vector represents the radial and poloidal wave numbers $\kappa$ and $m$, respectively.
Hence, ${\bf k} = (\kappa,m)$ and ${\bf k'}=(\kappa',m')$. \textcolor{black}{Here, $m'=nq(s_0)\pm \delta m$, where $q(s_0)$ is the q-value at $s=s_0$ and $\delta m$ decides the range of poloidal mode number.}
%For details of the trapped electron and trapped ion formulations the reader is referred to Ref.\cite{sb_tor}.
%
With the inclusion of the electromagnetic perturbations{\color{black}, but neglecting $\delta B_{||}$,} Eq.(1) above is modified as~\cite{gloria3,kbm4}
\begin{eqnarray*}
 {\tilde n}_{j}({\bf r};\omega) = - \left(\frac{q_{j}N}{T_{j}}\right)
\Bigg[ {\tilde\varphi} + \int d{\bf k} \exp\left({i {\bf k}\cdot{\bf r}}\right)\;
\int d{\bf v} \frac{f_{Mj}}{N}\left(\omega-\omega^{*}_{j}\right)\left(i
{\cal U}_{j}\right)
[\tilde \varphi({\bf k};\omega)-v_{\parallel}\tilde A_{\parallel}({\bf k};\omega)]J_{0}^{2}(x_{Lj})\label{eq_n}
\Bigg],
\end{eqnarray*}
where $\tilde A_{\parallel}$ is the parallel component of the vector potential. 
The perturbed parallel current density can be written as,
\begin{eqnarray*}
 {\tilde j_{\parallel j}}({\bf r};\omega) = - \left(\frac{q_{j}^2}{T_{j}}\right)
\Bigg[ \int d{\bf k} \exp\left({i {\bf k}\cdot{\bf r}}\right)\;
\int v_{\parallel}d{\bf v} f_{Mj}\left(\omega-\omega^{*}_{j}\right)\left(i
{\cal U}_{j}\right)
[\tilde \varphi({\bf k};)-v_{\parallel}\tilde A_{\parallel}({\bf k};)]J_{0}^{2}(x_{Lj})\label{eq_n}
\Bigg],
\end{eqnarray*}
With the quasi-neutrality condition Eq.(\ref{eq_closure}), and Ampere's law
$$\frac{1}{\mu_{0}}\nabla_{\perp}^2\tilde{A_{\parallel}} = -\sum_{j}\tilde{j_{\parallel j}}$$
We \textcolor{black}{finally} arrive at a linear system of equations as follows 
\begin{eqnarray}
\sum_{\bf k'}\sum_{j={\rm i,e}}{\cal \hat{M}}^{j}_{{\bf k},{\bf k'}}
      \begin{pmatrix}
        {\tilde \varphi}_{\bf k'} \\
        {\tilde A_{||\;\bf k'}} \\
      \end{pmatrix}=0
     \label{matrixEM_eq}
\end{eqnarray}     
This forms the core of the simulation model used in the present study. 
{\color{black}Note that Eq.(\ref{matrixEM_eq}) above represents a complex eigenvalue equation for the (complex) eigen frequency $\omega$, which is found numerically based on the Nyquist theorem. More details can be found in Ref. \cite{sb_thesis}}. \textcolor{black}{Moreover, a detail description of the set of equation included in Eq.(\ref{matrixEM_eq}) has been given in the Appendix. The geometry considered is circular, large aspect ratio, axisymmetric and unshifted. This code EM-GLOGYSTO is widely used to study many of the electrostatic and electromagnetic modes, such as the ITG~\cite{sb_tor, itg4}, TEM~\cite{tem5}, universal toroidal mode~\cite{unv2}, etc, in the electrostatic limit, and KBM~\cite{gloria3, kbm4, rg_prl}, and MTM~\cite{mtm6, mtm6a} modes in the electromagnetic limit. This code is capable of studying modes in the shorter wavelength as well and was already used to study the SWITG~\cite{sw8} with trapped electrons.}
%\vspace{-0.4cm}
\section{Results and Discussion}
\subsubsection{\bf{Parameters and Profiles}}
\noindent
For the present study, we consider the following profiles and parameters.\\
\begin{tabular}{lll}\label{Table1}
{\sf Parameters:} &    &{\sf Equilibrium Profiles:} \\ 
$\bullet$ B-field : $B_{0} = 1.0$ Tesla, \textcolor{black}{$\frac{m_i}{m_e}=1836$}  &   & $\bullet$ N-profile and T-profile           \\

$\bullet$ Temperature : $T_{0}=T(s_{0})=7.5\;keV$ &  &  $N(s)/N_{0} = \exp\left(-\frac{a\;\delta s_{n}}{L_{n0}}\;\tanh\left(\frac{s-s_{0}}{\delta s_{n}}\right)\right)$     \\
\textcolor{black}{$\bullet$ Density: $N_0=\beta(s)\times\left(B_0^2/2\mu_0\right)/(T_0^e+T_0^i)$ } &   & $T_{i,e}(s)/T_{0} = \exp\left(-\frac{a\;\delta s_{T}}{L_{T0}}\;\tanh\left(\frac{s-s_{0}}{\delta s_{T}}\right)\right)$\\

$\bullet$ Major Radius : $R = 2.0 \; m$ &   & $\delta s_{n}=0.35,\;\delta s_{T}=0.2$ at $s=s_{0}$    \\
$\bullet$ Minor Radius : $a = 0.5 \; m$ &     &  $\bullet$ q-profile and $\hat{s}$-profile:    \\
$\bullet$ \textcolor{black}{Radial coordinate} : $s=\rho/a,\; s_{0}=0.6$  &   &  $q(s) = 1.25 + 0.67\;s^{2} + 2.38\;s^3 - 0.06\;s^4$    \\
$\bullet$ $L_{n0}=0.2\;m,\;L_{T0}=0.08,\;m\rightarrow \eta_{i,e}(s_{0})=2.5$ &  &
     such that $q(s=s_{0})=2.0$; \\
$\bullet$ $\tau(s)=\frac{T_e(s)}{T_i(s)}=1$, $\epsilon_{n}=\frac{L_{n0}}{R}=0.1$, $\epsilon_{T}=\frac{L_{T0}}{R}=0.04$. &   &
     Shear ${\hat s}$ is positive and at $s=s_{0},\; {\hat s}=1$.\\
\end{tabular}\\

\begin{figure}[!h] %Figure 1
   \includegraphics[scale=0.16]{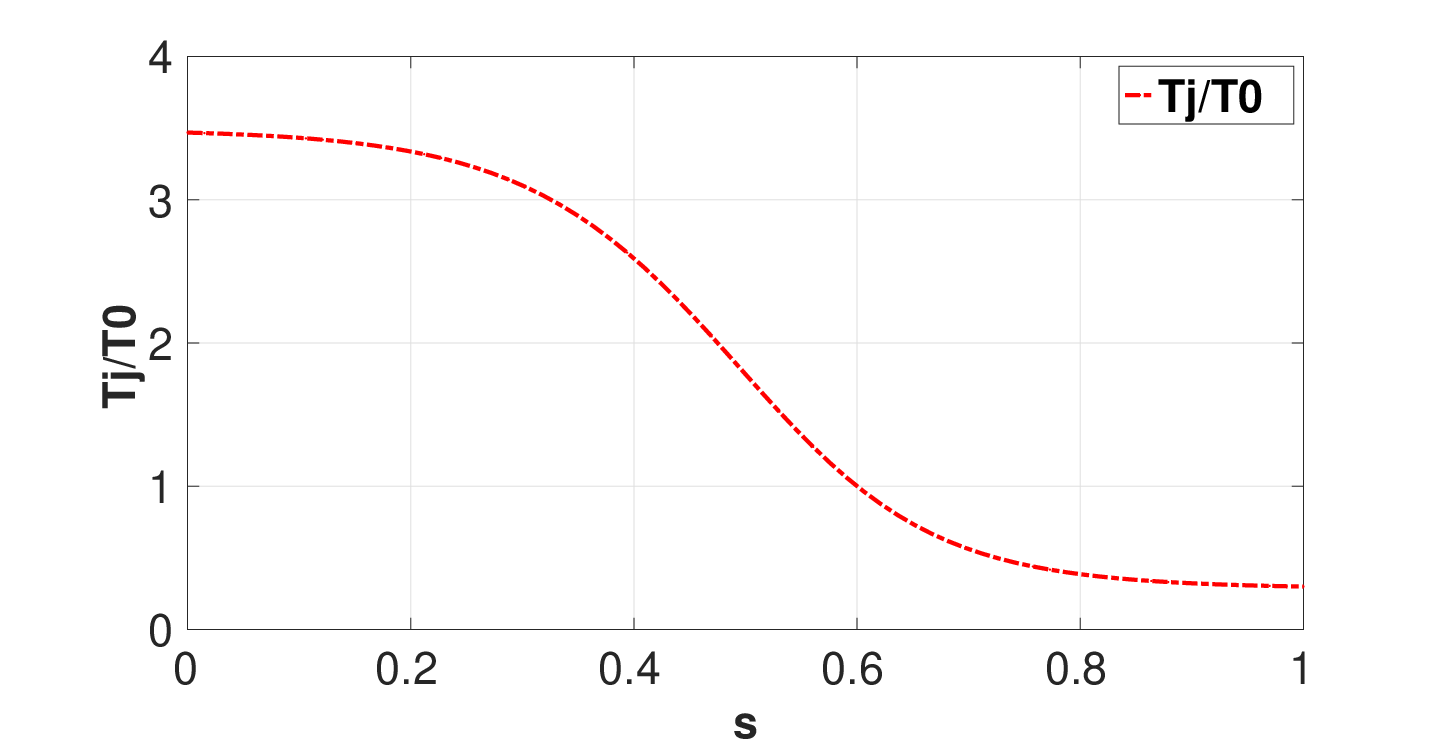}
    \includegraphics[scale=0.16]{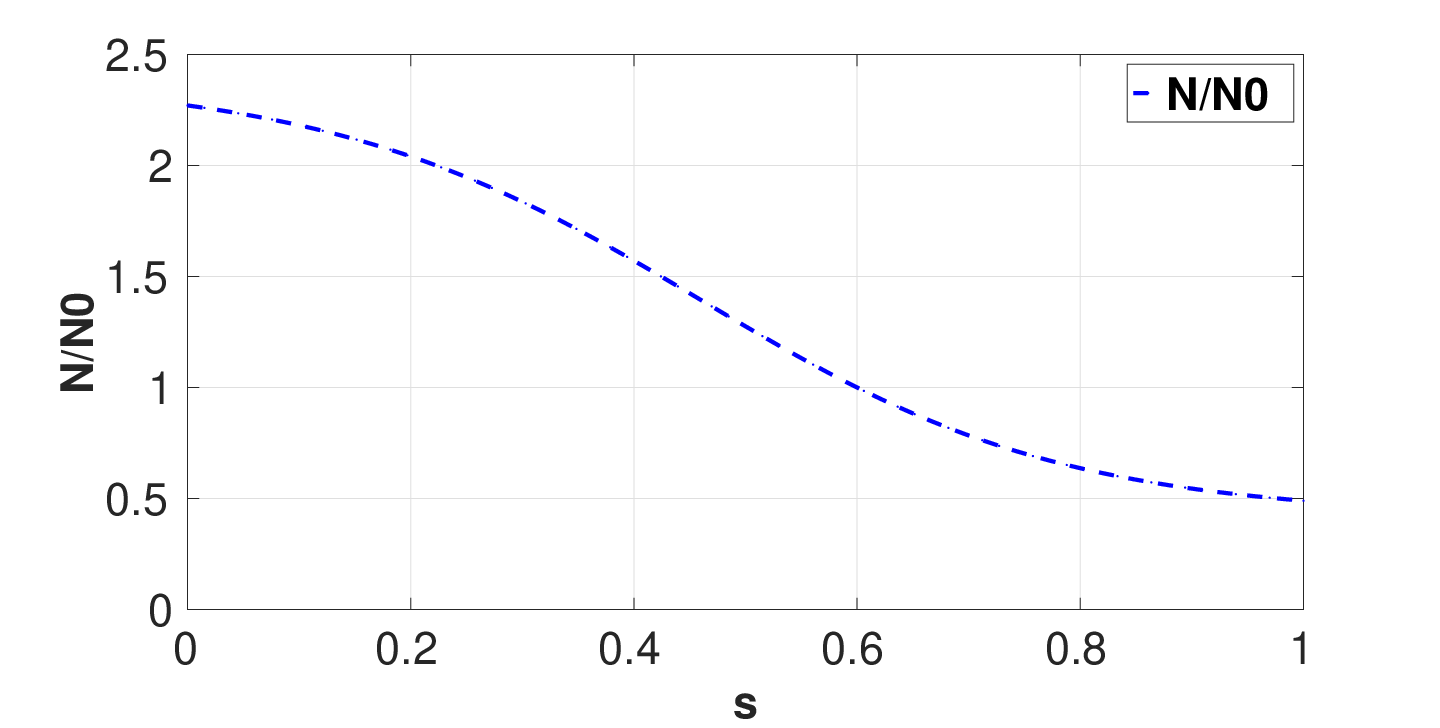}
\caption{Temperature profiles for electron and ion (left panel) and density profile (right panel). }
\label{fig1}
\end{figure}

\begin{figure}[!h] %Figure 1
    \includegraphics[scale=0.16]{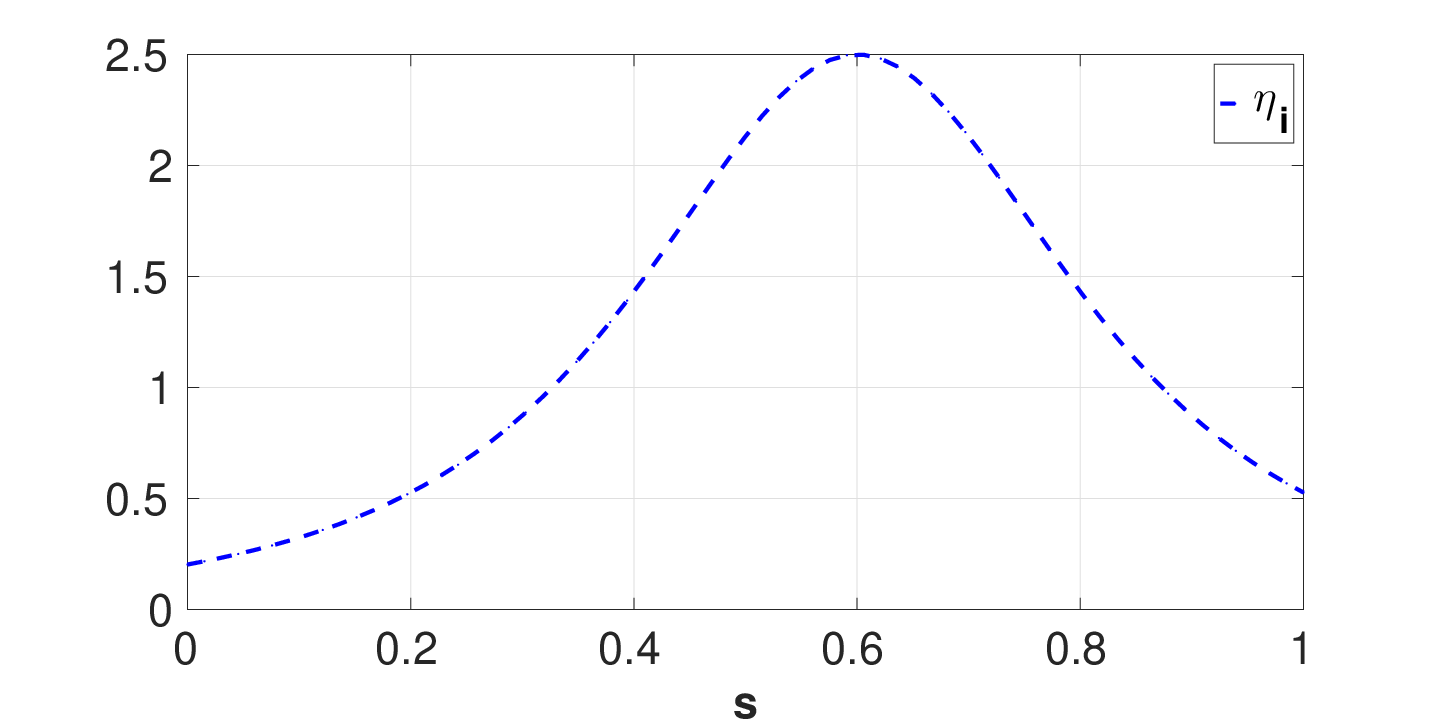}
    \includegraphics[scale=0.16]{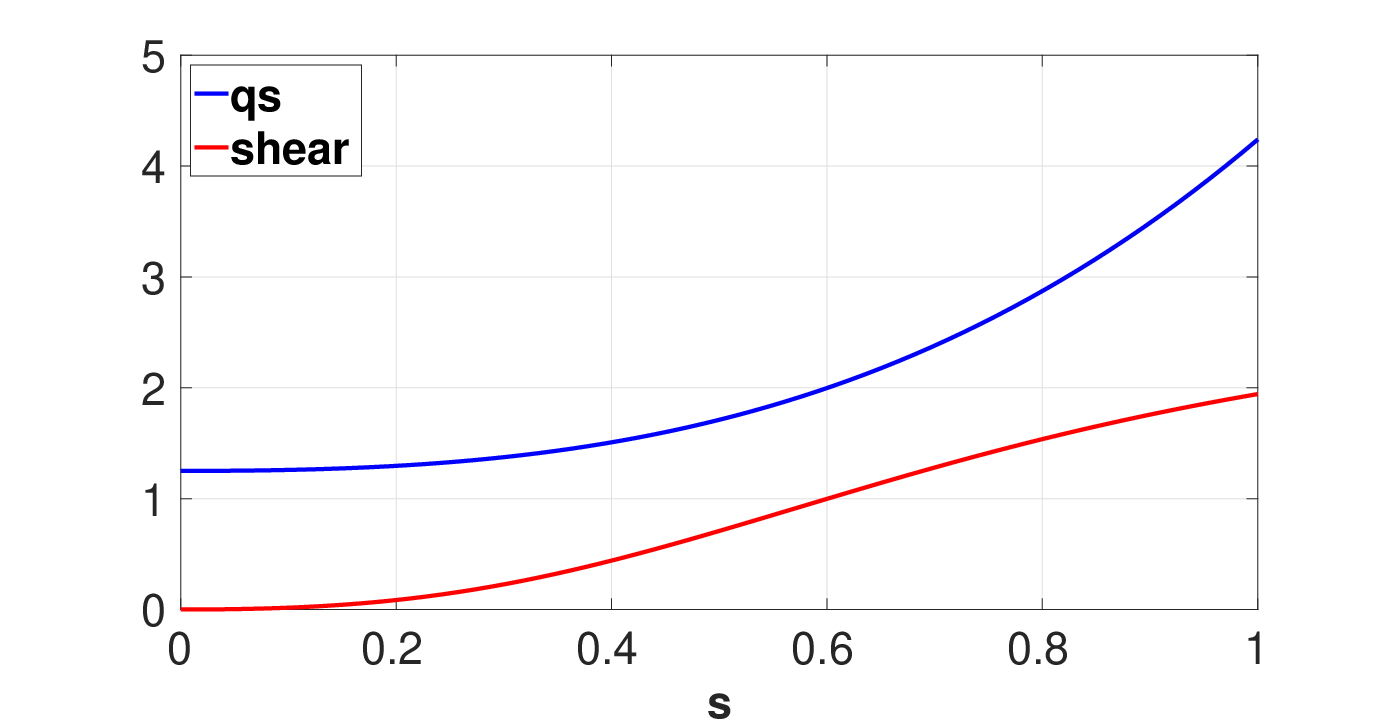}
\caption{$\eta_i$ profile (left panel) and safety factor and shear profiles (right panel). }
\label{fig2}
\end{figure}

\noindent
The corresponding profiles for density and temperature for ions and electrons are shown in Fig. \ref{fig1}. 
The left panel shows temperature while the right panel shows the densities for the species. \textcolor{black}{Note that the temperature and density profile are normalized by their respective equilibrium values {\it{i.e.}} $T_0$ and $N_0$. The value of $N_0$ can be calculated using the expression given in the Table 1 for Parameters and Profiles.}
The left panel of Fig. \ref{fig2} shows the $\eta_i$ profile. It is clear that it peaks at $s_0=0.6$. 
The $\eta_i$ profile is important, as this parameter determines the instability 
drive for the ITG and SWITG modes. The safety factor and  shear profiles are depicted in the right panel 
of Fig. \ref{fig2}. 
\subsubsection{\bf{Dispersion relation}}
\begin{figure}[!h] %Figure 3 
\includegraphics[scale=0.16]{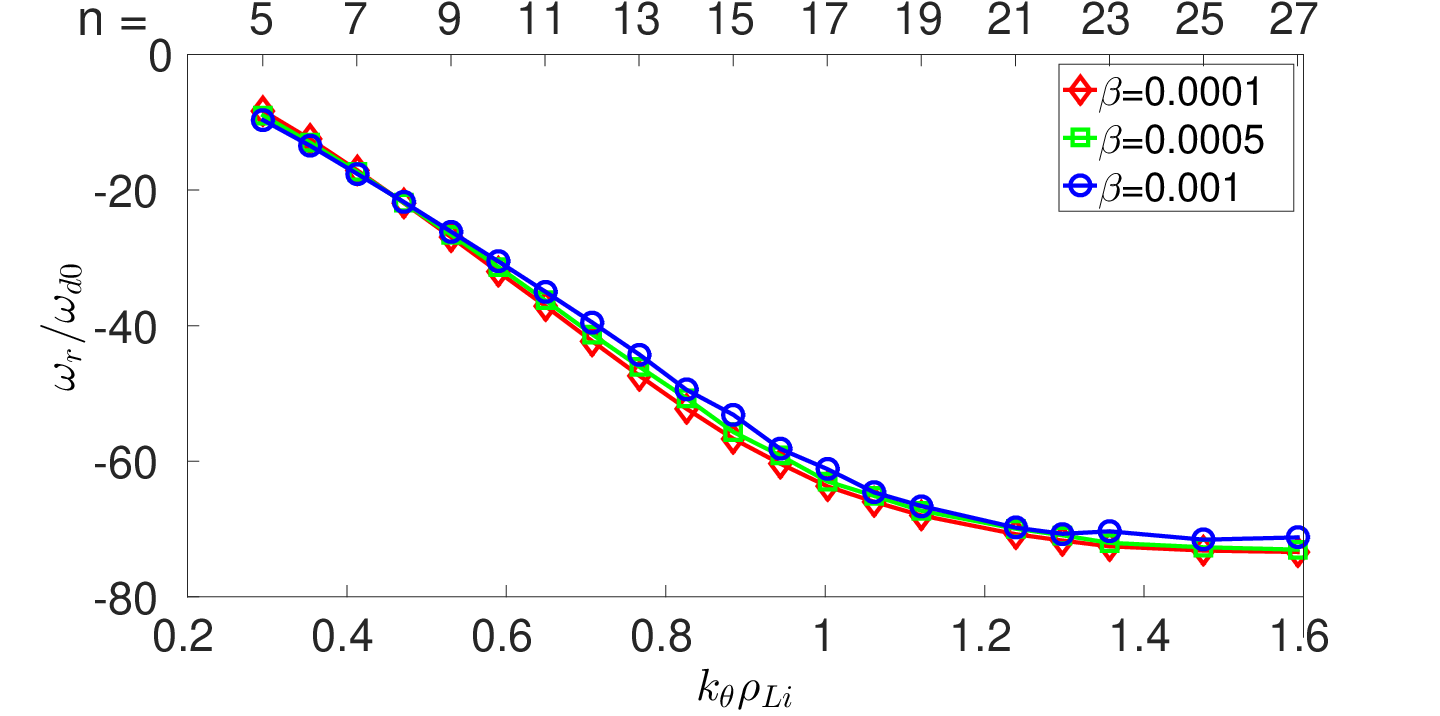}
\includegraphics[scale=0.16]{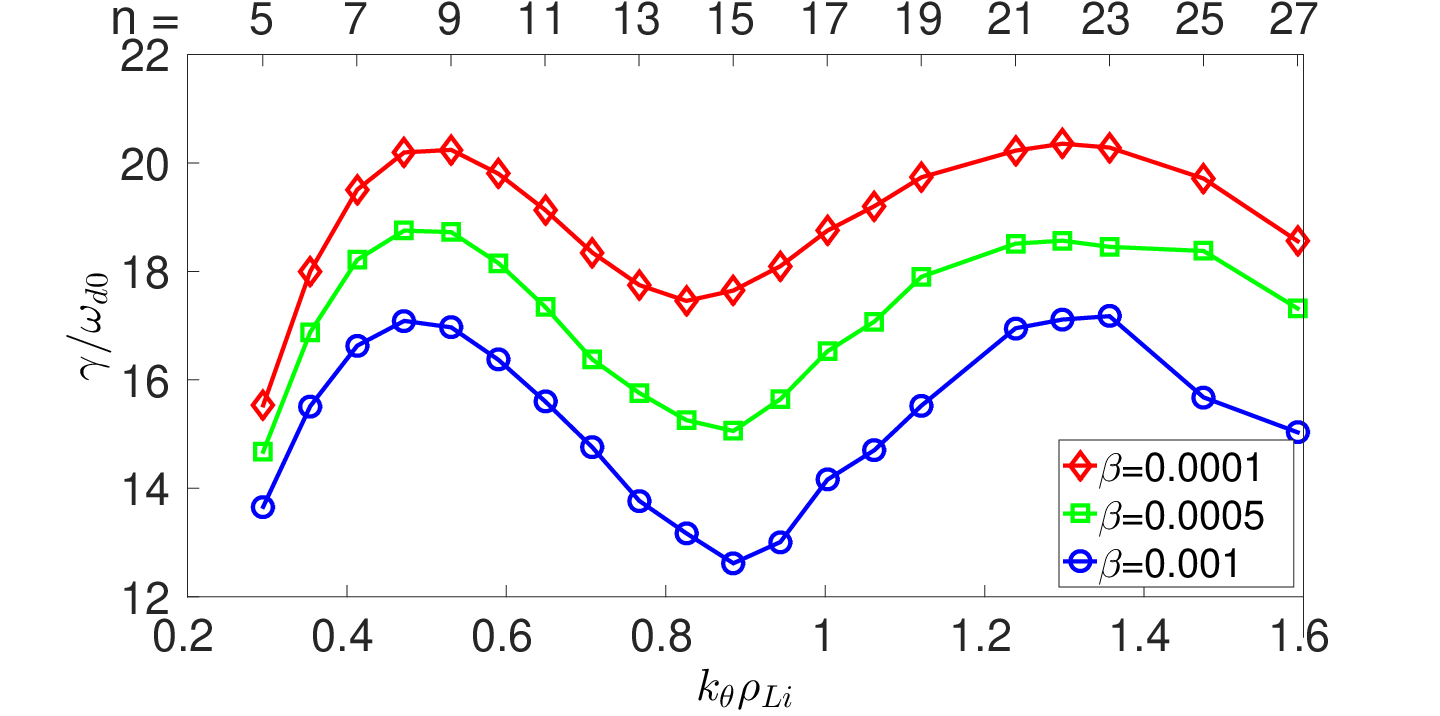}
\caption{Real frequency (left panel) and growth rate (right panel) \wrt \textcolor{black}{$k_\theta \rho_{Li}$ for different $\beta$ values. The upper axis shows the corresponding toroidal mode numbers.} The $\beta$ values considered are 0.0001 (red curve), 0.0005 $\simeq m_e/m_i$(green curve) and 0.001 (blue curve).}
\label{fig3}
\end{figure}

\begin{figure}[!h]
\includegraphics[scale=0.4]{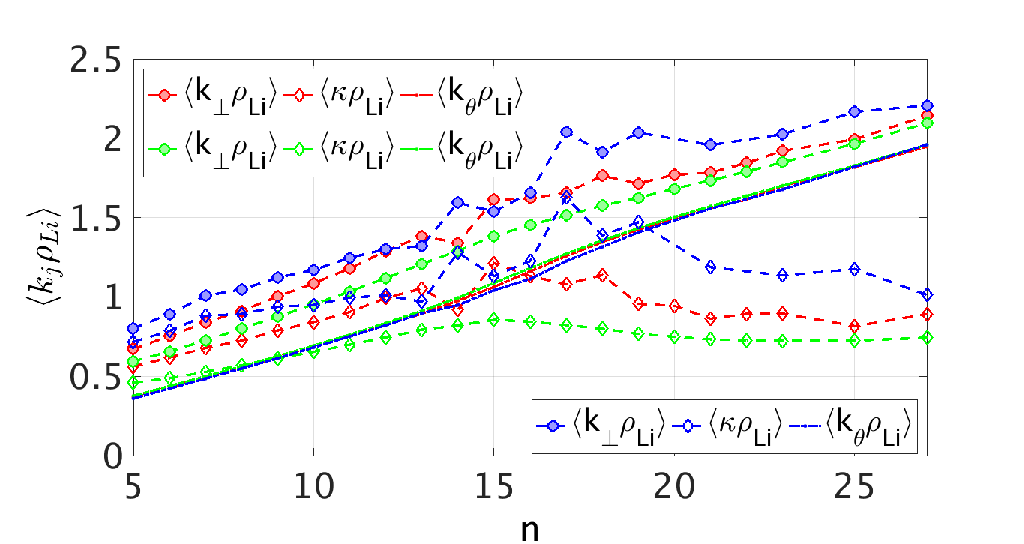}
\caption{Mode-averaged wavenumbers versus toroidal mode number $n$ for $\beta=0.0001$ (green color), $\beta=0.0005$ (red color), $\beta=0.001$ (blue color).}
\label{fig12}
\end{figure}

\noindent
In Fig. \ref{fig3} the real frequencies and growth rates are plotted for different values of the toroidal mode number n \textcolor{black}{and the corresponding $k_\theta \rho_{Li}$ values. Here the poloidal wave number $k_\theta=\frac{m}{r}=\frac{nq(s_0)}{s_0 a}$, where $m$ is the poloidal mode number}. \textcolor{black}{For all the simulation, poloidal mode number (m) has been varied from ($nq(s_0)-20$) to ($nq(s_0)+20$)}. 
The real frequency and growth rates are normalized by $\omega_{d0}=v_{thj}\rho_{Li}/a^2$.
These calculations
are carried out for different values of $\beta$, namely, 0.0001 (red curve), 0.0005 (green curve) and 0.001 (blue curve).
It is clear from the figure in the left panel that {\color{black}the} real frequency increases \wrt {\color{black}the} toroidal mode number $n$.
The real frequencies increase up to toroidal mode number $n\simeq20$. Beyond this point, the real frequencies stay 
virtually constant with $n$. This behavior is {\color{black}typical} of the SWITG mode \cite{sw6}. It is also evident that the real frequencies are weakly affected by the increase in $\beta$. The right panel of Fig. \ref{fig3} shows the corresponding growth rates 
with increasing toroidal mode number $n$. The growth rate increases initially \wrt $n$ and peaks around {\color{black}$k_\theta \rho_{Li}=0.5$} ($n\simeq 9)$ 
and then falls off with increasing $n$. This is the conventional ITG mode: with increasing toroidal 
mode number the resonance 
between mode frequency and magnetic drift enhances and therefore the growth rate increases. Beyond the peak, 
the resonance becomes weaker and the growth rate decreases. Also with the increase in the toroidal mode number, the finite Larmor 
radius effect comes in to play, because for a given ion Larmor radius with increasing toroidal mode number the wavelength 
decreases.  
However, with further increase in the toroidal mode number the real frequency turns to saturate and the growth 
rate again starts to increase. This is the regime of the SWITG mode. The growth rate increases up to {\color{black}$k_\theta \rho_{Li}=1.22$} ($n\simeq 21$) 
after which the growth rate again starts to decrease giving rise to the second hump which is the characteristic of 
the SWITG mode. The peak growth rates for both conventional ITG and SWITG mode are of similar strength. 
\noindent
Before discussing the \emf effects on the mode it is an appropriate place to revisit the 
theory of the SWITG mode following Ref.~\cite{sw6}. \textcolor{black}{The perturbed distribution function for ions in the local
limit $k_\perp \simeq k_\theta$ can be written as follows.
\begin{equation}\label{pert_fi}
f_i = - \frac{q_i f_{Mi}}{T_i} \tilde{\phi} + \frac{q_i f_{Mi}}{T_i}  \left( \frac{\omega - \omega_{*i}}{\omega - \omega_{di} - k_{||}v_{|}} \right) J^2_{0}(k_\perp \rho_{Li}) \tilde{\phi}
\end{equation}
The first term represents the adiabatic response while the second term represents the nonadiabatic response of the ions}. Upon integrating the perturbed distribution function \wrt velocity in the limit $\omega_{n}>\omega>(\omega_{di}+k_{\parallel}v_{\parallel})$, one can write the perturbed ion density in the electrostatic limit as follows~\cite{sw6}
\begin{eqnarray}
\tilde n_{i}=-\frac{q_{i}n_{o}}{T_{i}}\tilde \phi + \frac{q_{i}}{T_{i}}\tilde \phi\frac{\omega_{ni}(\eta_{i}/2-1)}
{\omega}I_{o}(k_{\perp}^2\rho_{Li}^2)exp(-k_{\perp}^2\rho_{Li}^2),
\end{eqnarray}
where $I_{o}$ stands for the zeroth order modified Bessel function and $\omega_{ni}=-(v_{thi}/L_{n}) (k_{\theta}\rho_{Li})$.
Then applying the  quasi-neutrality condition with adiabatic electrons one arrives at the following relation 
\begin{eqnarray}
\omega=\Bigg( \frac{\tau}{\tau+1}\Bigg)\Bigg (\frac{\eta_{i}}{2}-1 \Bigg)\omega_{ni}I_{o}(k_{\perp}^2\rho_{Li}^2)exp(-k_{\perp}^2\rho_{Li}^2),
\label{dispersion1}
\end{eqnarray}
The monotonic increase in the mode frequency with toroidal mode number at lower $n$ region (conventional ITG) and then saturation
of the mode frequency at higher $n$ (SWITG) are easily understood from the above relation using the properties of the 
scaled modified Bessel function $I_{o}(b)e^{-b}\rightarrow 1/\sqrt(2\pi b)$ for large values of $b$. For small $k_{\perp}^2\rho_{Li}^2$, the mode frequency $\omega$ behaves as $k_{\perp}\rho_{Li}$ and for larger
$k_{\perp}^2\rho_{Li}^2$ it remains virtually constant. 
The second hump in the growth rate appears as a result of a second resonance between \textcolor{black}{the} toroidal magnetic drift term 
$\omega_{di}$ of the ions with the mode frequency which is constant at higher toroidal mode number. 
It is also clear from the Fig. \ref{fig3} that the SWITG also suffers from finite Larmor radius 
stabilization. This can be understood from 
the expression of the  nonadiabatic part in Eq.\ref{pert_fi}. At very high $n$ or equivalently, high 
$k_{\perp}\rho_{Li}$, the ion magnetic drift frequency $\omega_{di}$ leads the mode frequency $\omega$.
The nonadiabatic part of the perturbed ion density, in the limit $\omega_{di}>>\omega$ and large $k_{\perp}^2\rho_{Li}^2$ 
will then decrease as $\frac{R}{L_{n}}I_{o}(k_{\perp}^2\rho_{Li}^2)exp(-k_{\perp}^2\rho_{Li}^2)$. For 
greater detail of these calculations, readers are referred to Ref~\cite{sw6}. It is evident from Fig. \ref{fig3} that, although the real frequencies are not very much affected by the increase 
in the value of $\beta$, the growth rates of both conventional ITG and SWITG mode suffer a substantial reduction in 
magnitude. The reduction in the growth rates of the conventional ITG mode is well known. There have been 
many linear and nonlinear studies reporting the stabilization of the ITG mode by the electromagnetic 
perturbations. However, the role of electromagnetic perturbations on the SWITG mode is perhaps hitherto not investigated in detail using global gyrokinetic model. 
Since the advanced operating regimes  where SWITG mode can be unstable also have higher $\beta$ values, 
it is important to explore the \emf effect on the mode.
Thus this present study shows that the SWITG mode suffers stabilization in the presence of the electromagnetic 
perturbation and one has to take in to account this effect as well when studying the ion heat flux 
induced by SWITG in real experiments. In the presence of the \emf perturbation, the mode 
couples with the Alfven perturbations. The field lines are thus bent in the presence of $\beta$~\cite{gloria3,wlnd}. 
This field line bending by 
the electromagnetic perturbation is therefore responsible for the stabilization of the SWITG mode. \textcolor{black}{The stabilization by field line bending may be
understood looking at the energy equation for stability in the ideal MHD theory. If $\delta W$ is the change in the potential energy under a small perturbation, the condition $\delta W < 0$ leads to an unstable situation. If we retain only field line bending and pressure driven terms for simplicity, then $\delta W$ can be written as $\delta W = \int dv \left[ \frac{|\tilde{B}|^2}{2\mu_0} - \left( \xi_\perp \cdot \nabla p \right) \left(\kappa \cdot \xi_\perp ^{*} \right) \right]$, where $\tilde{B}$, $p$, $\xi_\perp$ and $\kappa$ are perturbed magnetic field, pressure, plasma displacement and curvature, respectively. The first term which comes from field line bending is always positive and therefore provides stabilization}. The mode averaged wavenumbers versus toroidal mode number $n$ is shown in Fig. \ref{fig12}. It is clear from the figure that with the increase in $n$, the value of $\langle \kappa \rho_{Li}\rangle$ increases initially up to $n\simeq15$, but with further increase in the value of $n$, SWITG branch starts (see Fig. \ref{fig3}) and $\langle \kappa \rho_{Li}\rangle$ tends to saturates. Similar to the ITG branch, for SWITG branch as well, the contribution of $\langle \kappa \rho_{Li}\rangle$ to $\langle k_\perp \rho_{Li}\rangle$ is significant. It is also important to note that with the increase in $\beta$ values, $\langle k_\theta \rho_{Li}\rangle$ value does not change much, but $\langle \kappa \rho_{Li}\rangle$ increases. Hence, for higher $\beta$ values, the relative contribution of $\langle \kappa \rho_{Li}\rangle$ to $\langle k_\perp \rho_{Li}\rangle$ also increases. For example, for $n=21$,  $\langle k_\theta \rho_{Li} \rangle = 1.571, 1.560, 1.556$ and  $\langle \kappa \rho_{Li}\rangle$/$\langle k_\perp \rho_{Li}\rangle$ = 0.446, 0.484, 0.606 for respective values of $\beta=0.0001,~0.005,~0.001$. This suggests the radial length scale becomes shorter and shorter for higher $\beta$ values and also comparable to the poloidal length scale. This strong radial behavior shows the kinetic as well as global nature of the SWITG mode. 
\subsection{Mode structures for ITG and SWITG}
\begin{figure}[!h] %Figure 4 
\includegraphics[scale=0.24]{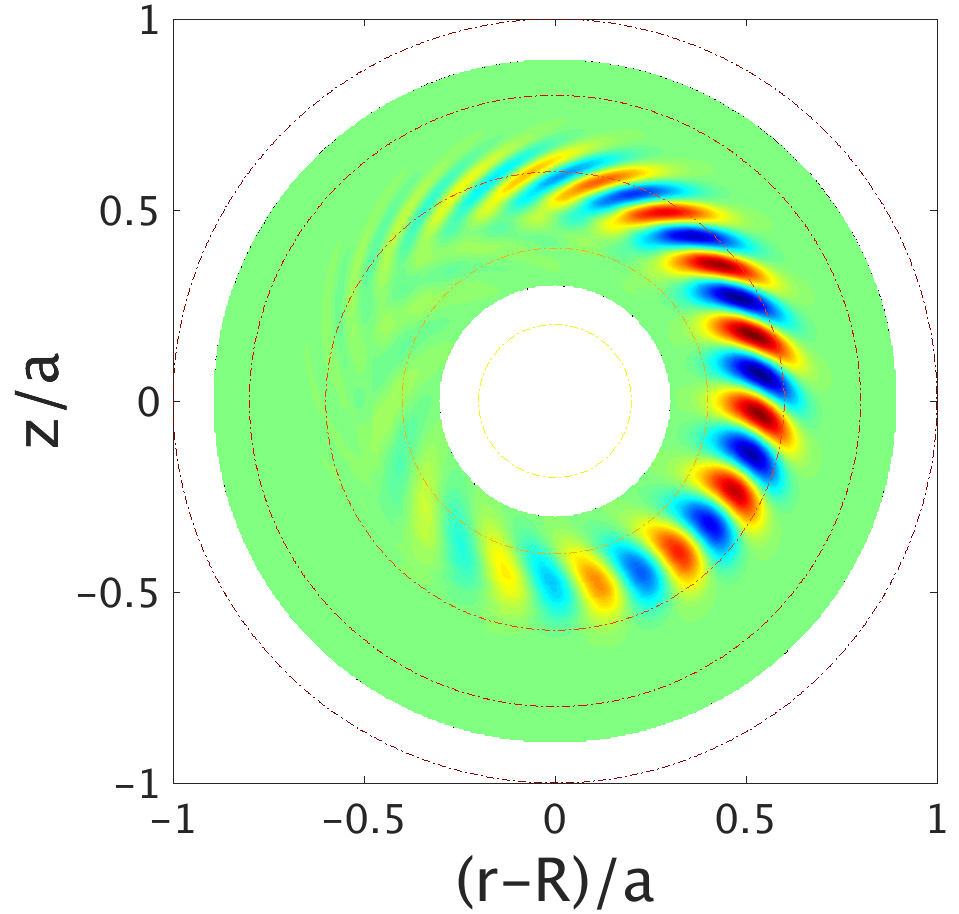}
\includegraphics[scale=0.24]{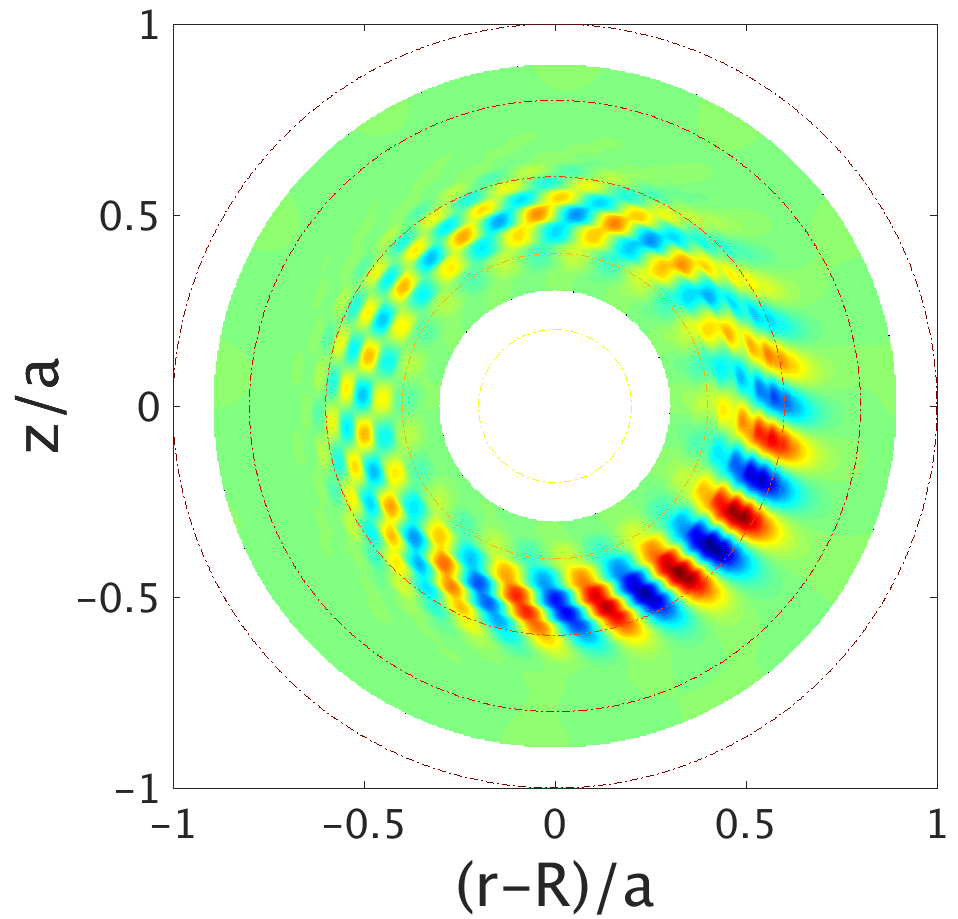}
\caption{Mode structure of $\phi$ (left panel) and $A_{\parallel}$ (right panel) for $n=9$ corresponding to the ITG 
mode $\beta=0.0001$.}
\label{fig4}
\end{figure}

\begin{figure}[!h] %Figure 5 
\includegraphics[scale=0.24]{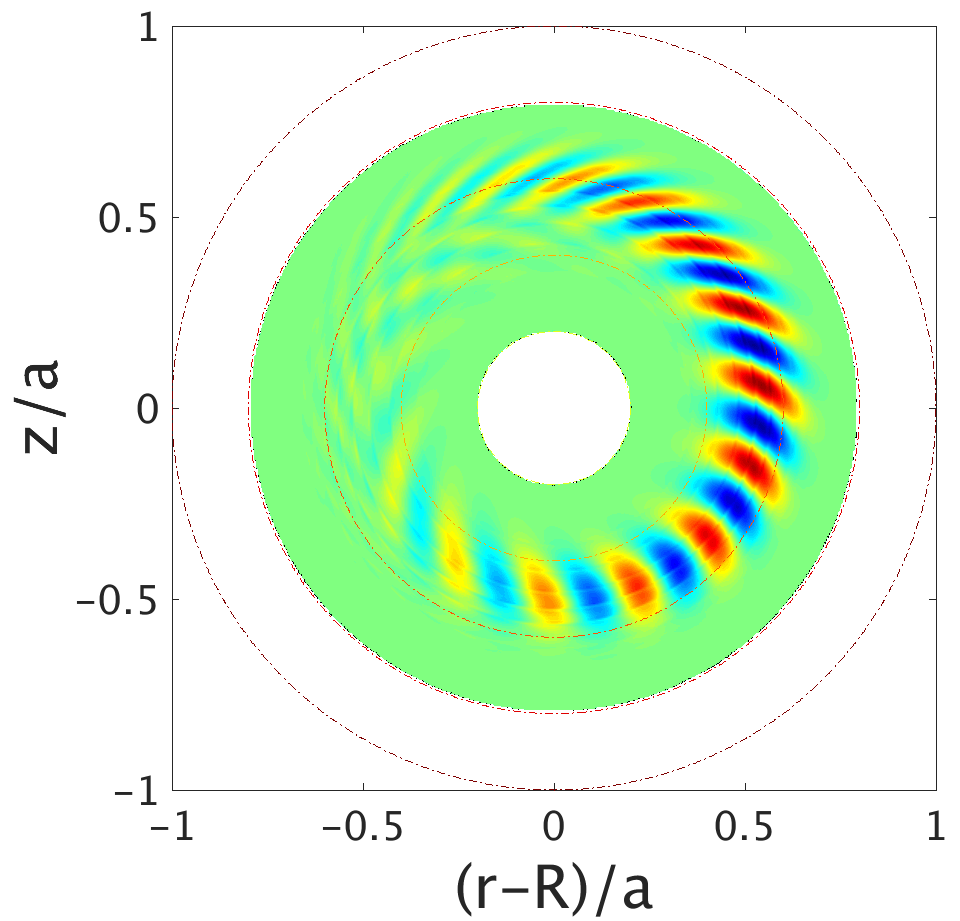}
\includegraphics[scale=0.24]{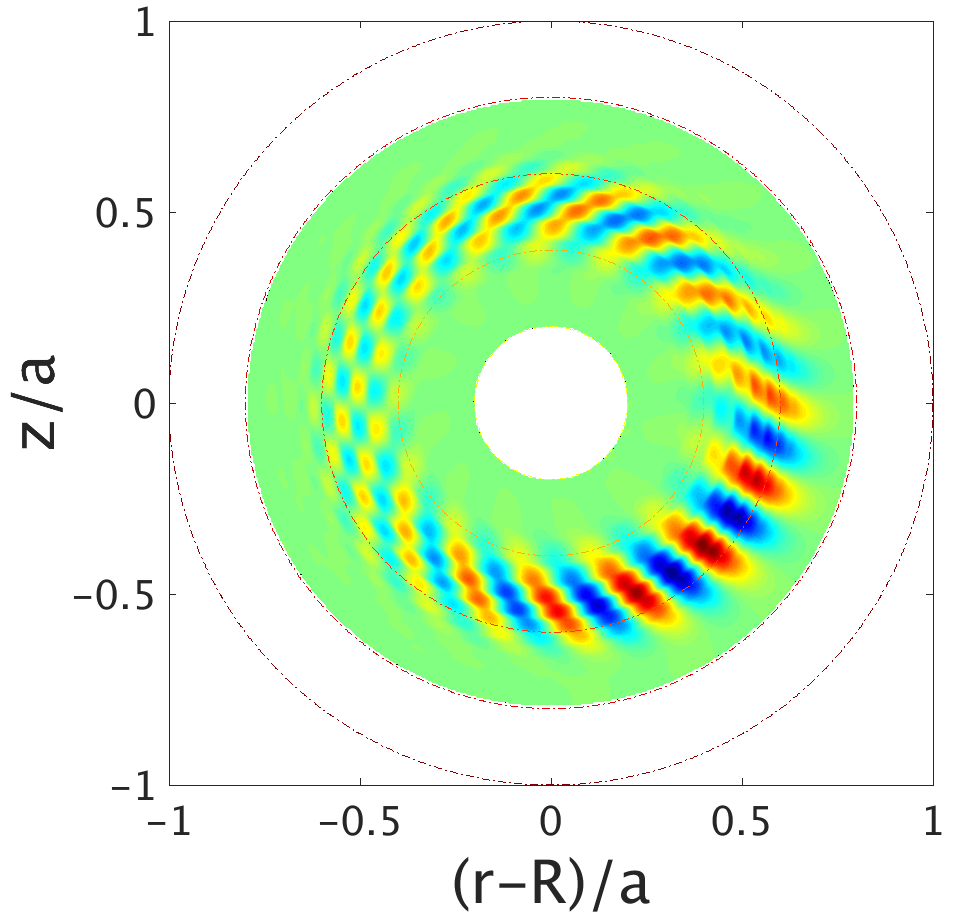}
\caption{Mode structure of $\phi$ (left panel) and $A_{\parallel}$ (right panel) for $n=9$ corresponding to the ITG mode $\beta=0.001$.}
\label{fig5}
\end{figure}

\begin{figure}[!h] %Figure 6 
\includegraphics[scale=0.24]{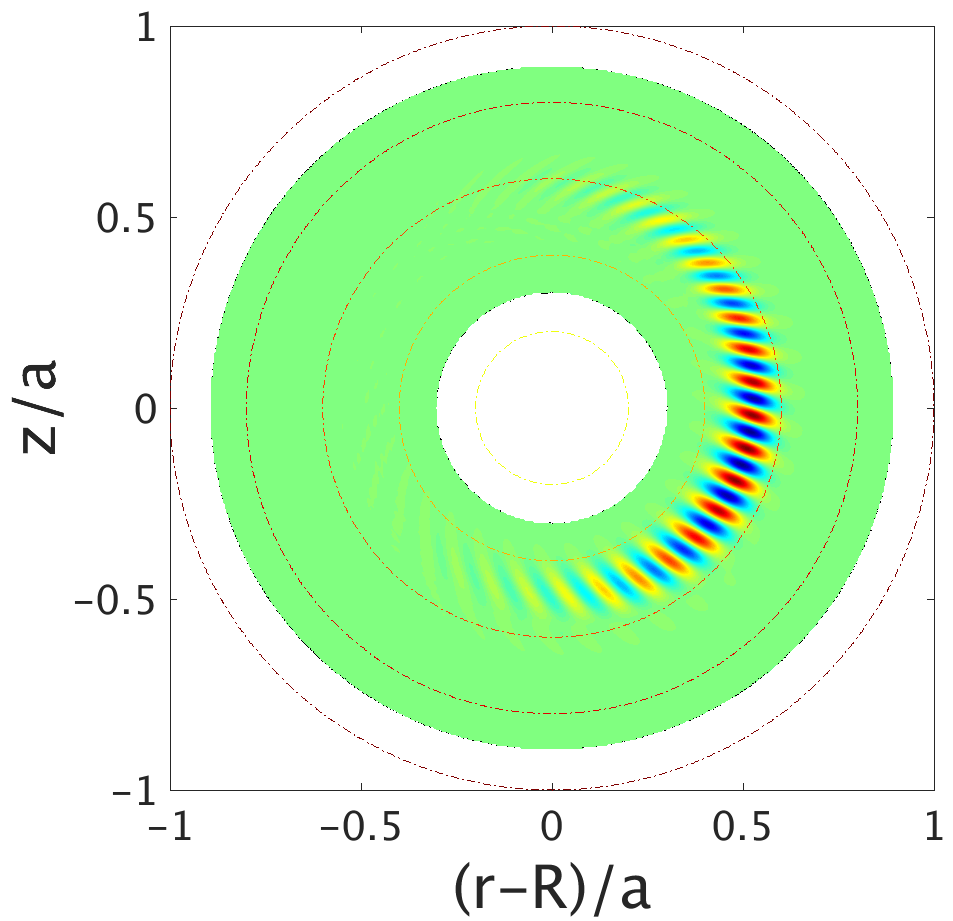}
\includegraphics[scale=0.24]{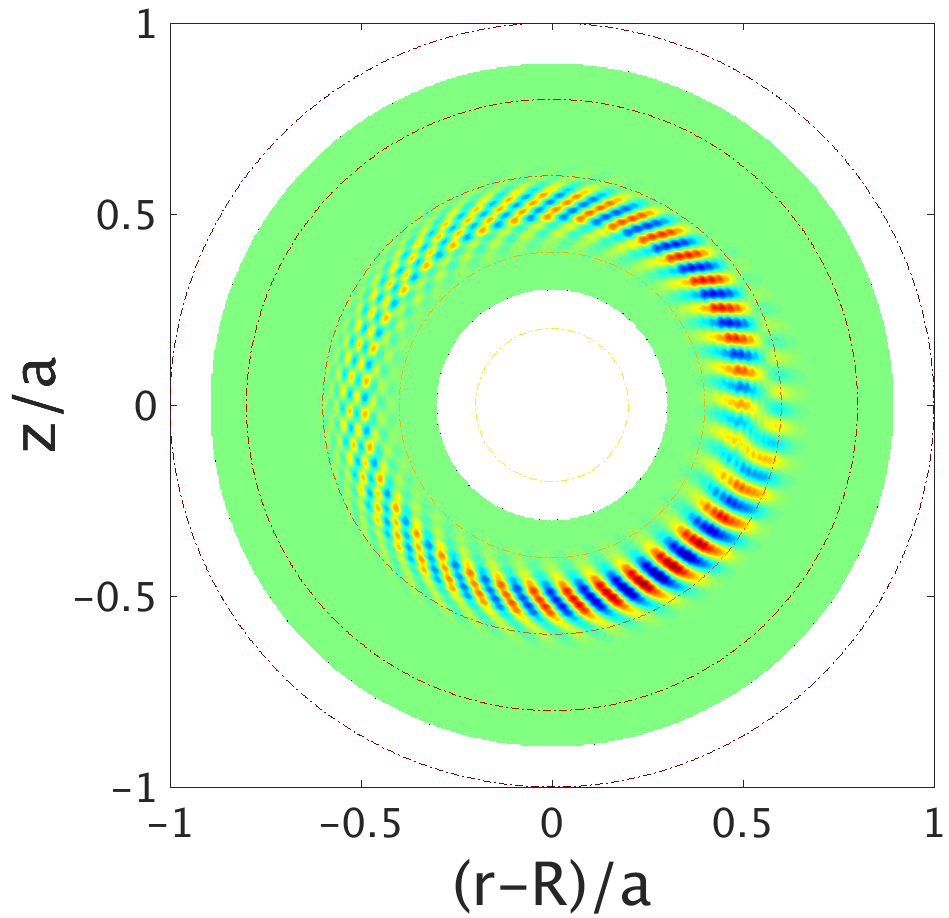}
\caption{Mode structure of $\phi$ (left panel) and $A_{\parallel}$ (right panel) for $n=21$ corresponding to the SWITG mode for $\beta=0.0001$.}
\label{fig6}
\end{figure}

\begin{figure}[!h] %Figure 7 
\includegraphics[scale=0.24]{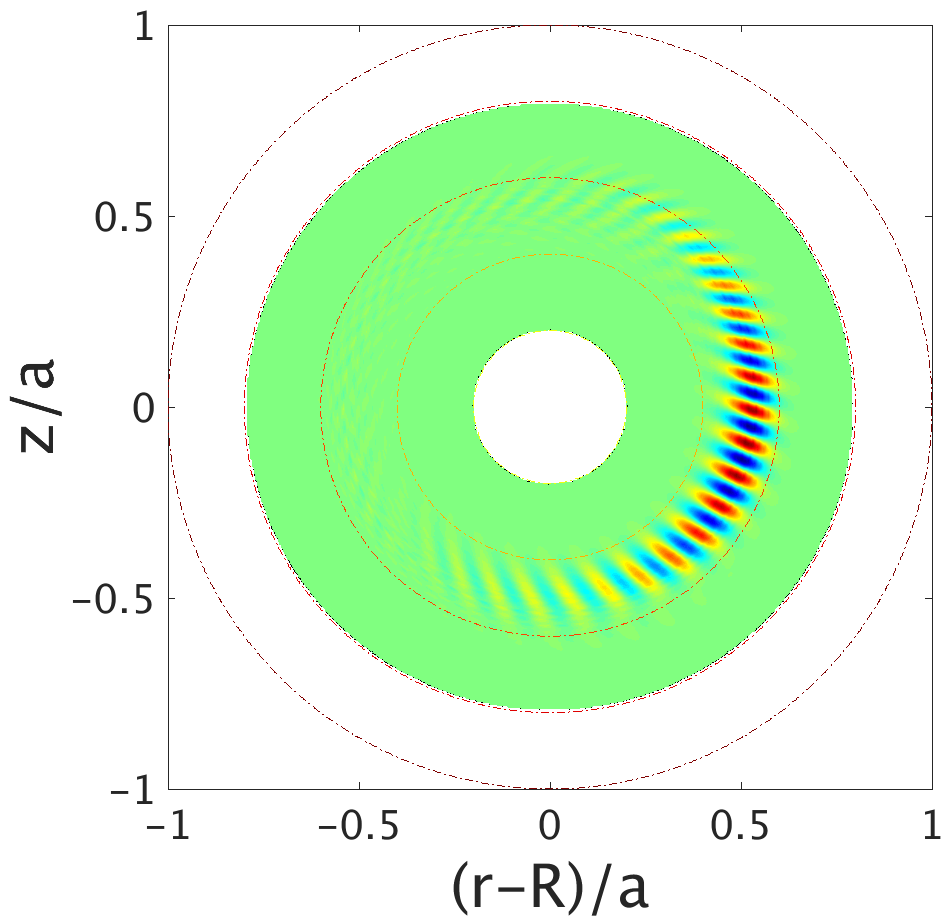}
\includegraphics[scale=0.24]{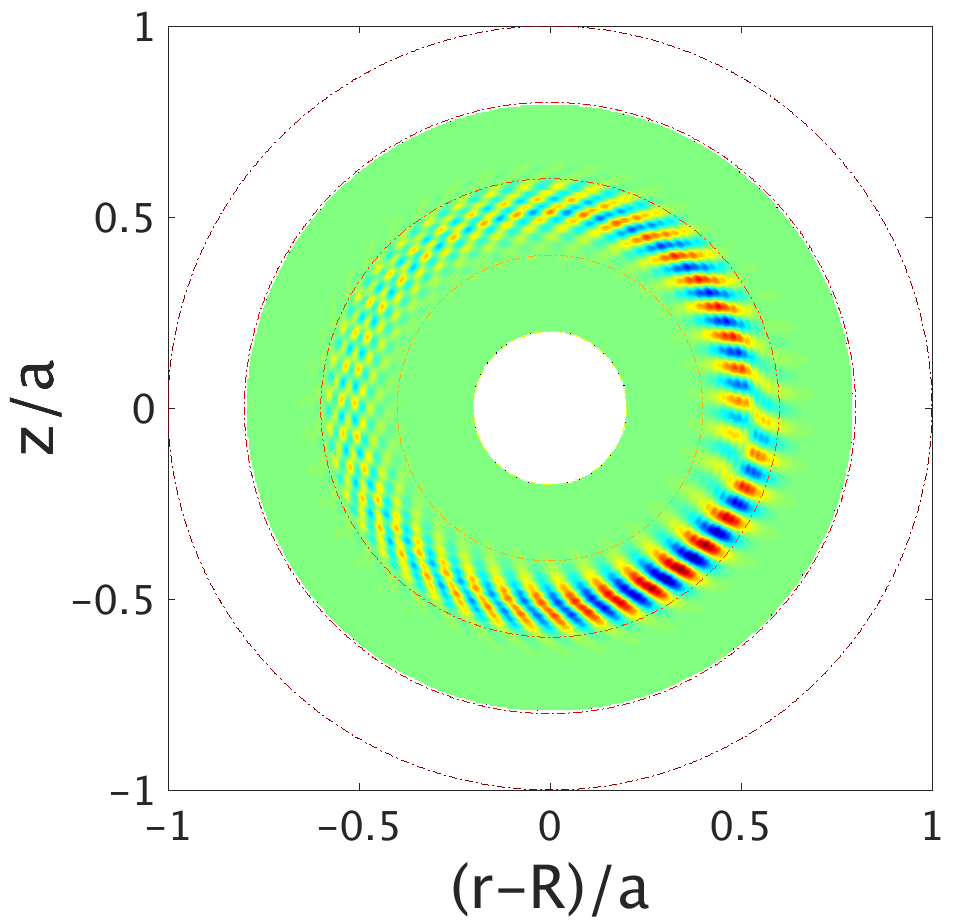}
\caption{Mode structure of $\phi$ (left panel) and $A_{\parallel}$ (right panel) for $n=21$ corresponding to the SWITG mode for $\beta=0.001$.}
\label{fig7}
\end{figure}

\noindent
Fig. \ref{fig4} shows the mode structures for electrostatic and electromagnetic potentials
$\phi$ and $A_{\parallel}$ for $\beta=0.0001$. The toroidal mode number corresponding 
to the mode structure is $n=9$. It is clear from the figure that 
for both $\phi$ and $A_{\parallel}$ the mode structures exhibit ballooning character.
Similarly, Fig. \ref{fig5} depicts the mode structure for $\phi$ and $A_{\parallel}$ for 
$\beta=0.001$ for toroidal mode $n=9$. Similar to the modes in Fig. \ref{fig4}, the modes exhibit ballooning 
structure which is a characteristic of ITG mode. 
Figs. \ref{fig6} and \ref{fig7} show the mode structures of $\phi$ and $A_{\parallel}$ for SWITG mode for $\beta=0.0001$ and 
$\beta=0.001$, respectively. The toroidal mode number for these figures 
is $n=21$.
\textcolor{black}{The figures clearly show that the {\color{black}SWITG} mode is ballooning in nature and more 
localized both radially and poloidally {\color{black}as compared to the normal ITG mode}. This observation suggests that the
mode might be susceptible to stabilization due to toroidicity~\cite{sw8}.}
\subsection{$\beta$ scan}
\begin{figure}[!h] %Figure 8 
\includegraphics[scale=.16]{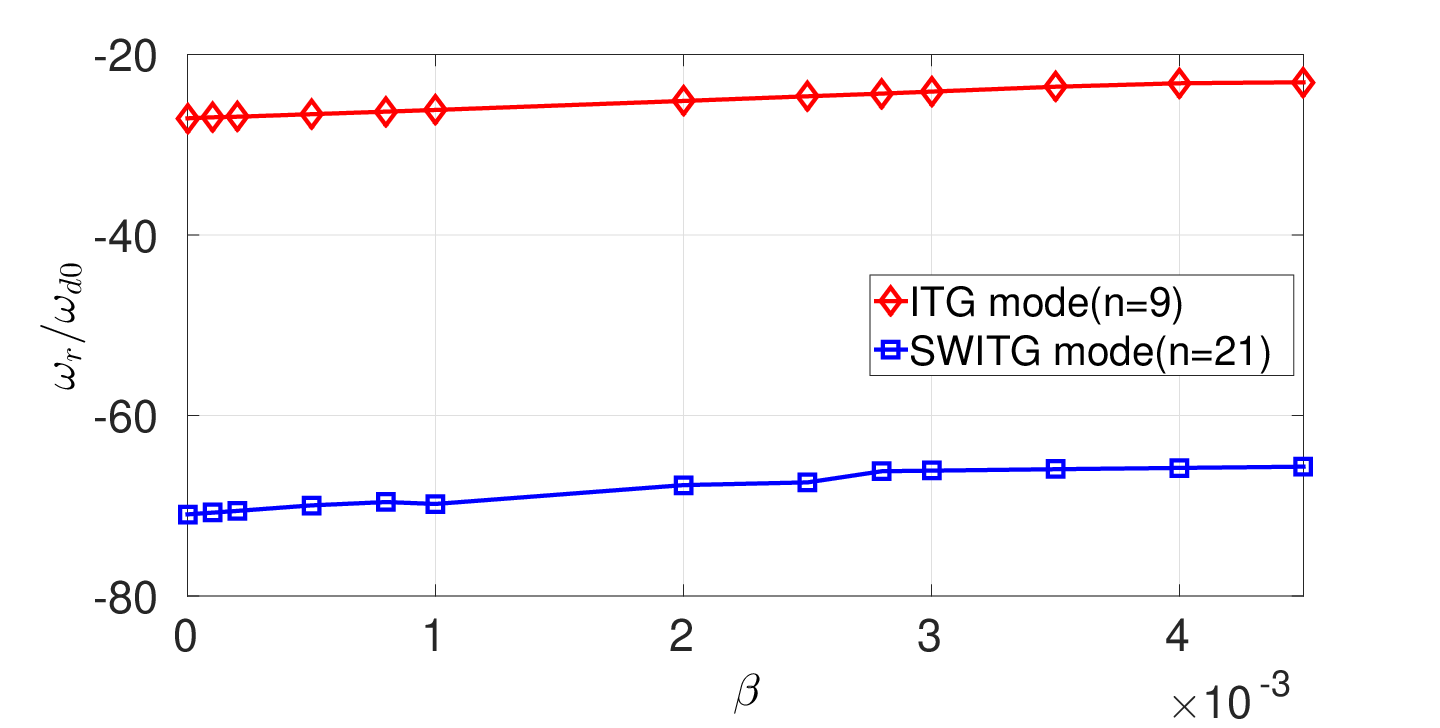}
\includegraphics[scale=.16]{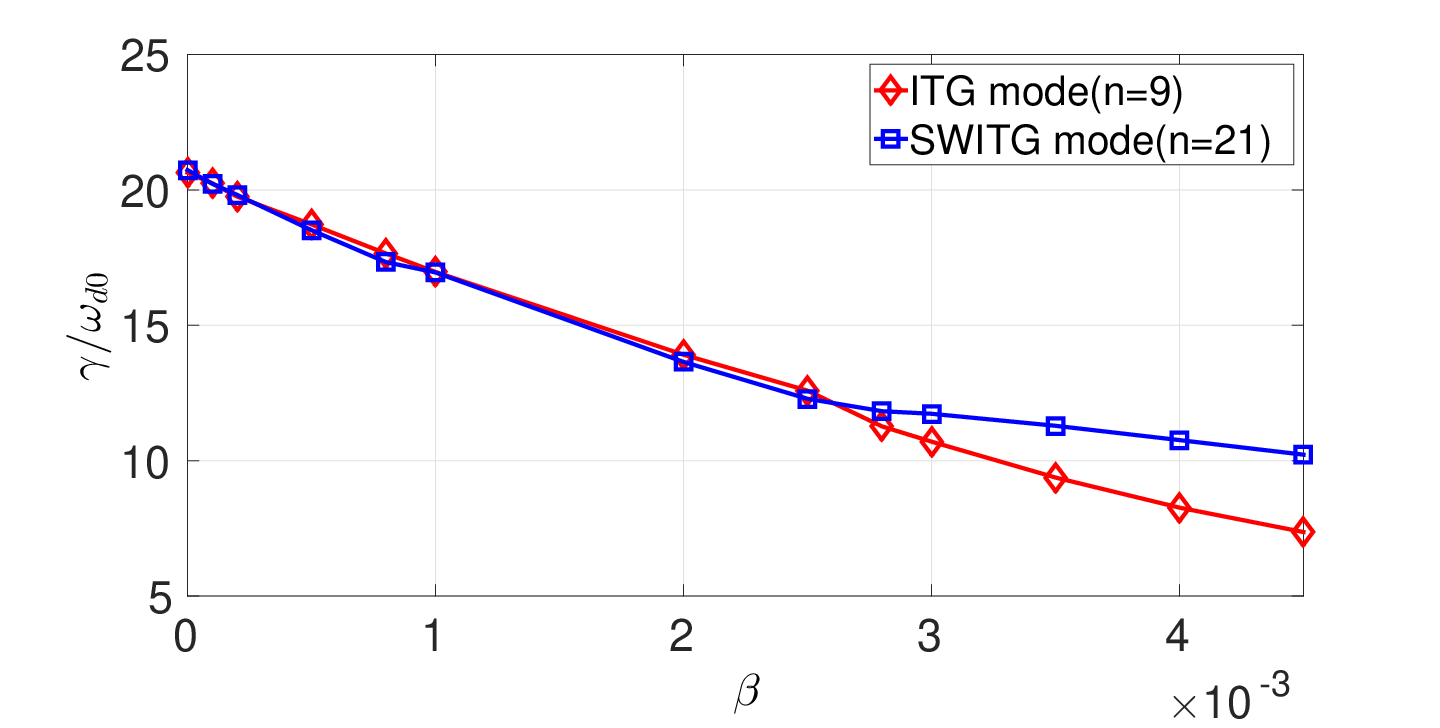}
\caption{Real frequency (left panel) and growth rates (right panel) for the ITG and SWITG mode \wrt $\beta$.}
\label{fig8}
\end{figure}

\begin{figure}[!h] %Figure 9 
\includegraphics[scale=.3]{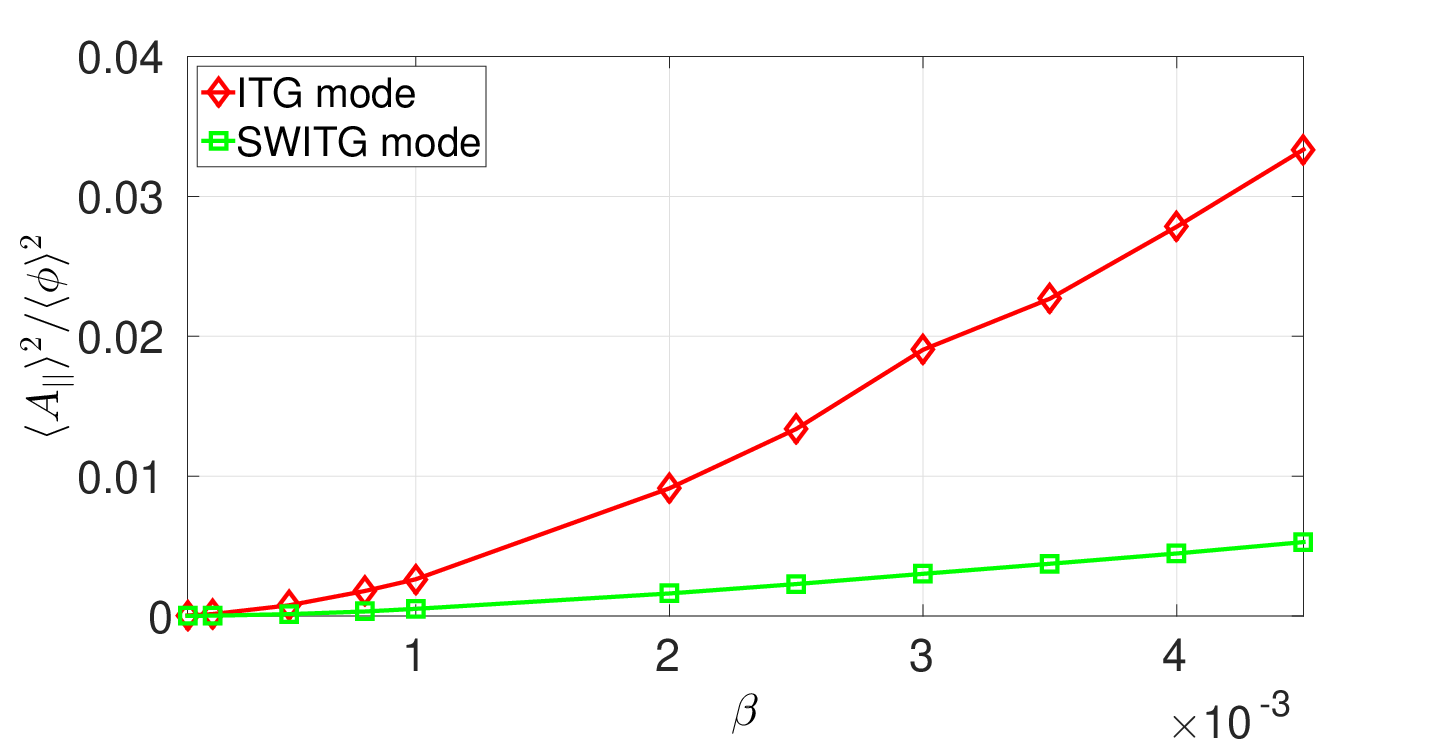}
\caption{Ratio of $\langle A_{||}\rangle ^2/\langle\phi\rangle^2$ \wrt $\beta$ for the ITG and SWITG mode.}
\label{fig9}
\end{figure}

\noindent
In this section, we investigate the effect of $\beta$ on the real frequency and growth rates
of SWITG mode and compare them with the conventional ITG mode. The real frequencies and growth rates 
are calculated for increasing values of $\beta$ and are shown in Fig. \ref{fig8}. The real frequencies versus 
$\beta$ are shown in the left panel and the growth rates versus $\beta$ are shown in the right panel. 
It is clear from the figure that with increasing $\beta$ 
the real frequency increases weakly with $\beta$. In contrast to this, the growth 
rate decreases with increasing $\beta$. {\color{black}However, at a higher $\beta$ growth rate of SWITG mode decreases slowly with respect to $\beta$ compared to that of ITG mode.}
Thus we can conclude that the SWITG mode suffers  $\beta$ stabilization like its long wavelength counterpart. The relative strength of the electromagnetic to the 
electrostatic character is shown in Fig. \ref{fig9}. This is expressed as the ratio
${\langle A_{\parallel} \rangle}^2/{\langle \phi \rangle}^2$. It is clear
from the figure that the ratio increases almost linearly {\color{black}with increasing $\beta$}.
It is also observed that
the value of the ${\langle A_{\parallel} \rangle}^2/{\langle \phi \rangle}^2$ ratio for SWITG is lower
compared to that of the conventional ITG mode. \textcolor{black}{Here, the value of $m_i/m_e$ is 1836.}
\subsection{$\eta_i$ scan}
\begin{figure}[!h] % Figure 10
\includegraphics[scale=.16]{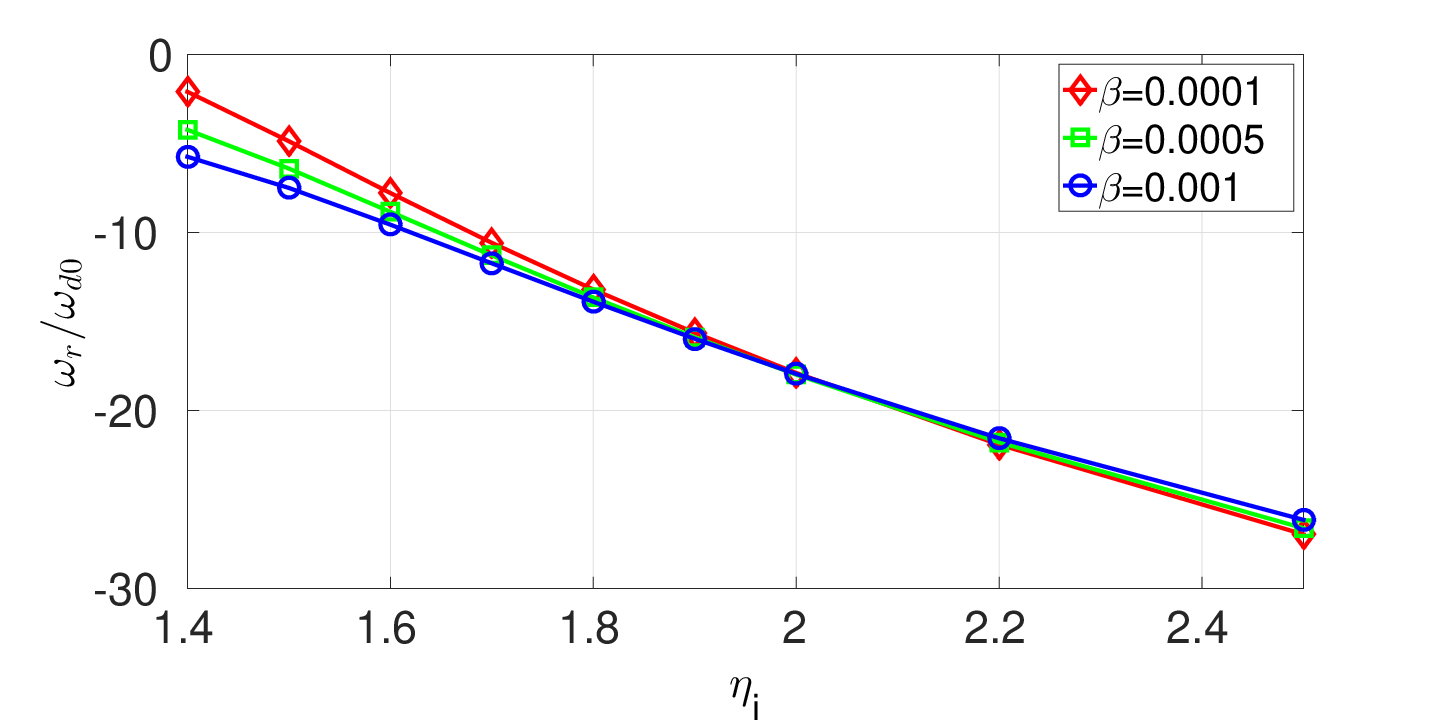}
\includegraphics[scale=.16]{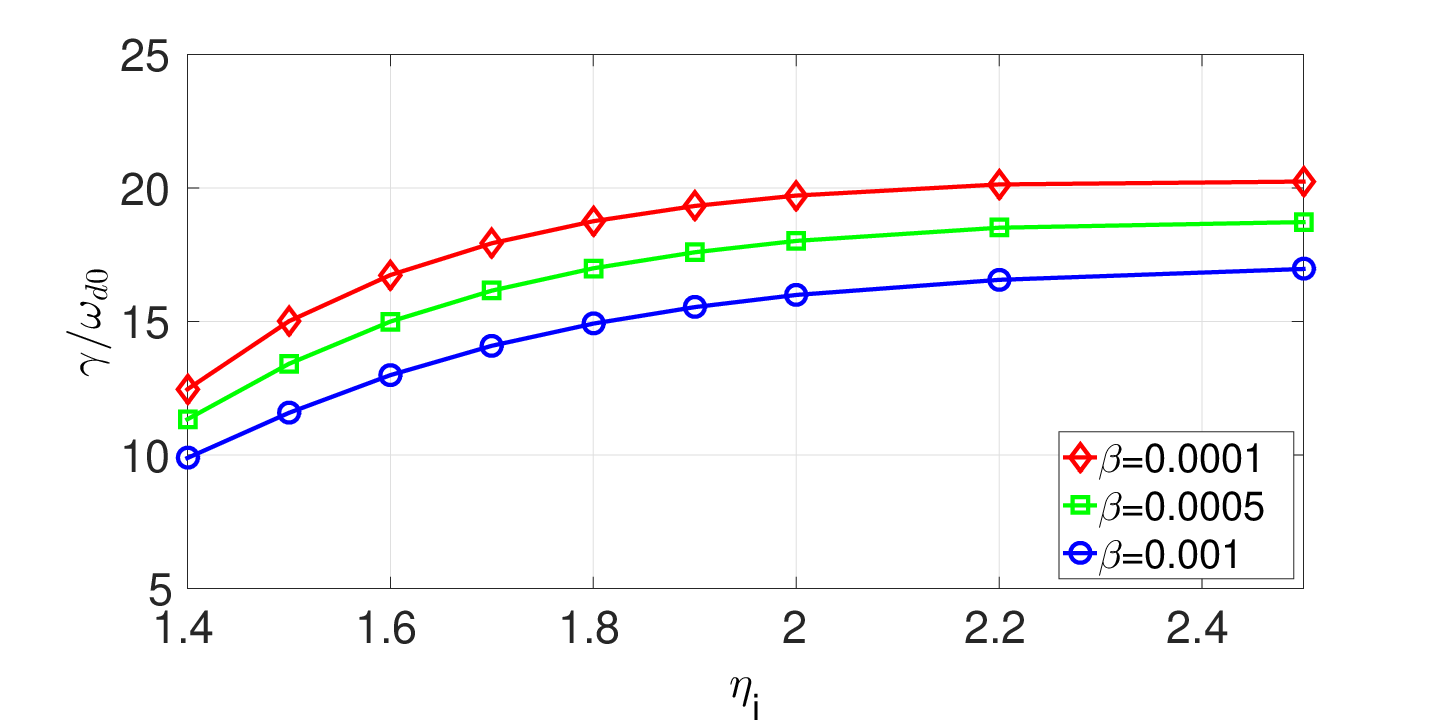}
\caption{Real frequency (left panel) and growth rates (right panel) \wrt $\eta_i$ for the ITG mode.} 
\label{fig10}
\end{figure}

\begin{figure}[!h] % Figure 11
\includegraphics[scale=0.16]{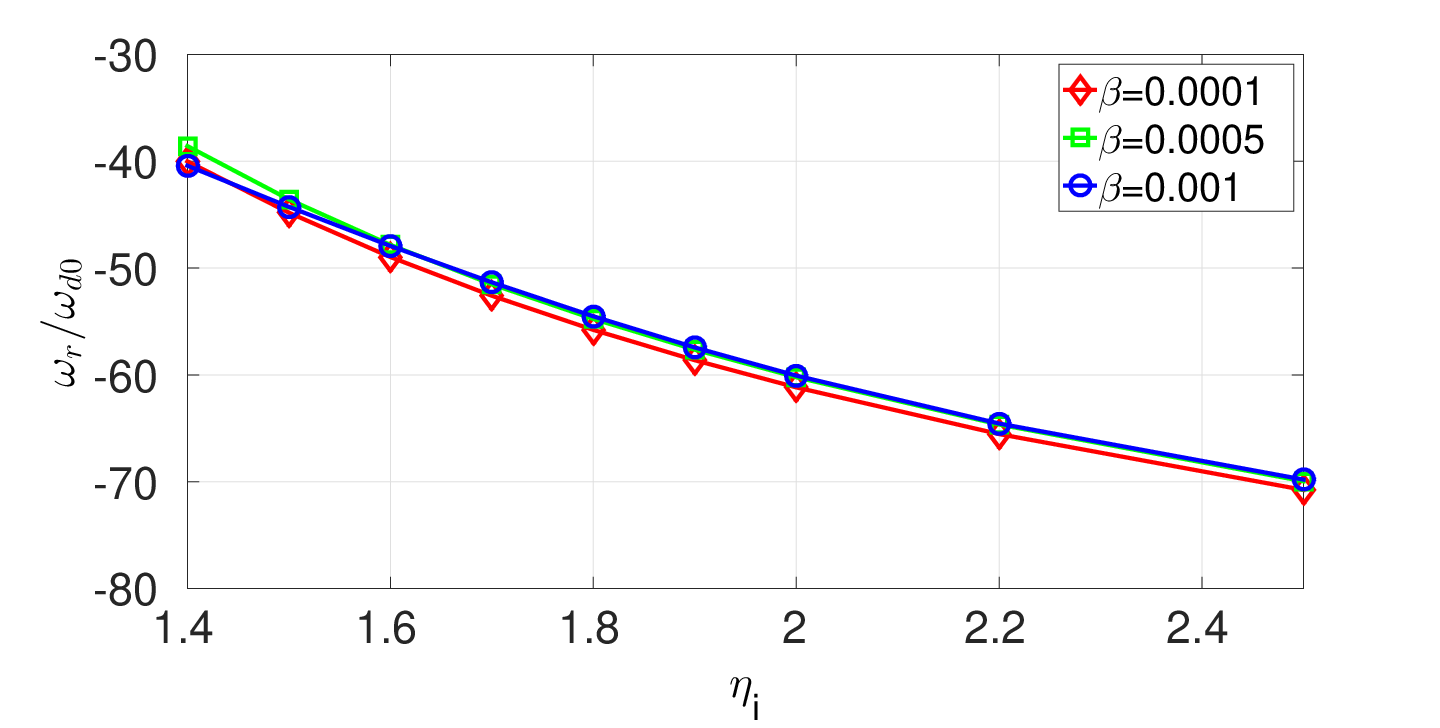}
\includegraphics[scale=0.16]{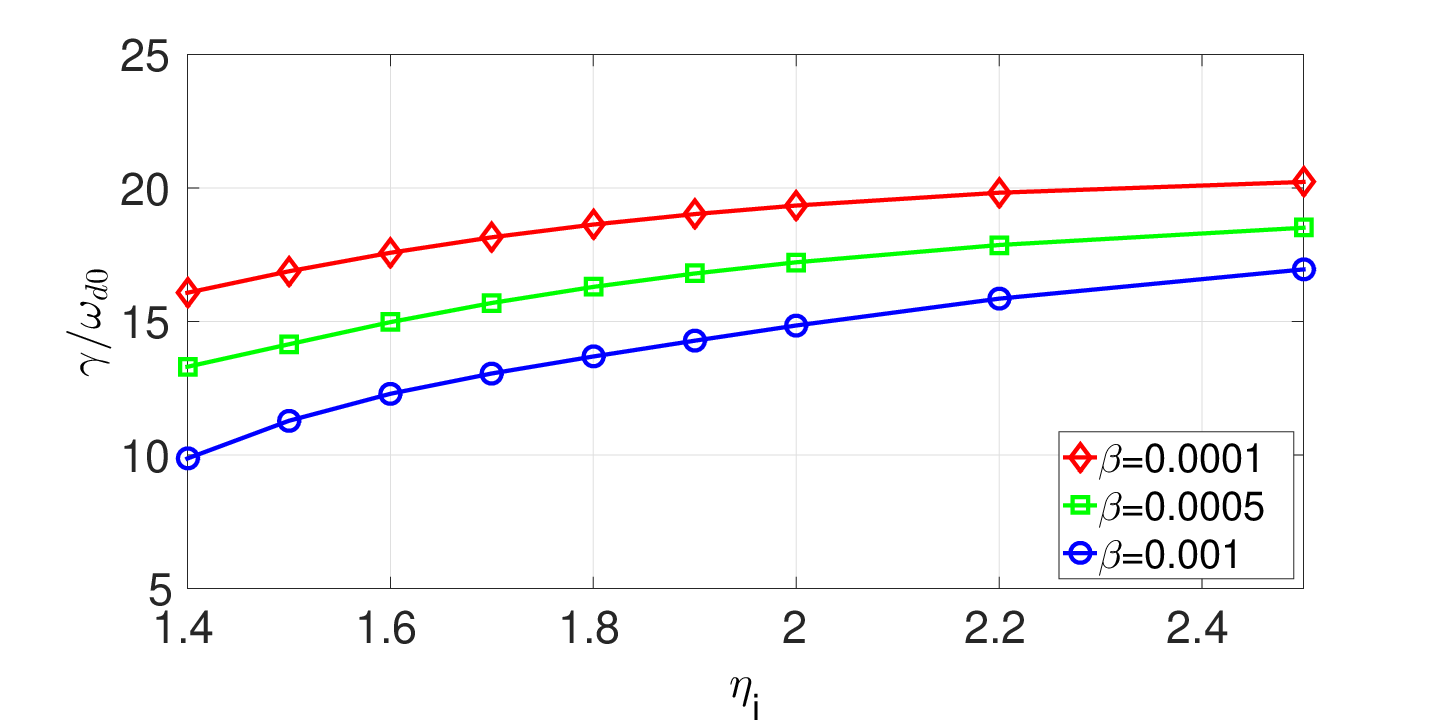}
\caption{Real frequency (left panel) and growth rates (right panel) \wrt $\eta_i$ for the SWITG mode.} 
\label{fig11}
\end{figure}

\noindent
For temperature gradient modes, such as the SWITG mode, $\eta_i$, which is the ratio of density and temperature 
gradient scale lengths, is a very important parameter. When $\eta_i$ is above 
a certain value the SWITG mode becomes unstable. Beyond the threshold value, 
the growth rate increases monotonically with increasing $\eta_i$. To explore this physics in 
the presence of \emf perturbation,
we calculate the mode frequency and growth rate for increasing values of $\eta_i$ for both the 
ITG and the SWITG modes for different values of $\beta$. The results for the ITG mode are shown in Fig. \ref{fig10} and the corresponding 
results for the SWITG mode are shown in Fig. \ref{fig11}. 

\indent
\textcolor{black}{The growth rates increase with $\eta_i$ in both the cases. An explanation for the $\eta_i$ dependence of the ITG and SWITG mode can be given following Ref.\cite{wlnd}. The free energy for ITG mode and SWITG mode comes from the temperature gradient. In the presence of density gradient, simultaneous convection in temperature and density gradients takes place resulting in a competition between convection and expansion in the energy equation. During the outward convection, plasma moves from a higher density region to a lower density region where expansion takes place. At the same time, plasma moves from a higher temperature region to a lower temperature region increasing the temperature. The expansion leads to the cooling of the plasma that competes with the increase in the temperature. For a displacement $\xi$, the change in the temperature can be written as~\cite{wlnd}, $\delta T = - \xi \cdot \nabla T + \alpha \xi \cdot \nabla n$, where $\alpha$ is a coefficient giving the cooling due to expansion. This relation clearly shows that for $\delta T$ to be positive, $\eta$ has to exceed a certain threshold value. The net free energy available depends upon the relative strength of the density gradient and temperature gradient~\cite{wlnd}. This is manifested in the dependence on $\eta_i$}. Thus with increasing $\eta_i$ the free energy available 
to render the mode unstable also increases. This leads to an increase in the 
growth rate. This explains the increase in the growth rates with increasing $\eta_i$ 
as observed in Figs. \ref{fig10} and \ref{fig11}. It is also evident from the Figs. \ref{fig10} and \ref{fig11} that the real frequency
increases with $\eta_i$. The real frequency is proportional to the 
diamagnetic frequency which is proportional to the gradient. That is 
why the real frequency increases with increasing $\eta_i$. 
\subsection{Mixing length calculation of flux}
\begin{figure}[!h] % Figure 13
\includegraphics[scale=0.2]{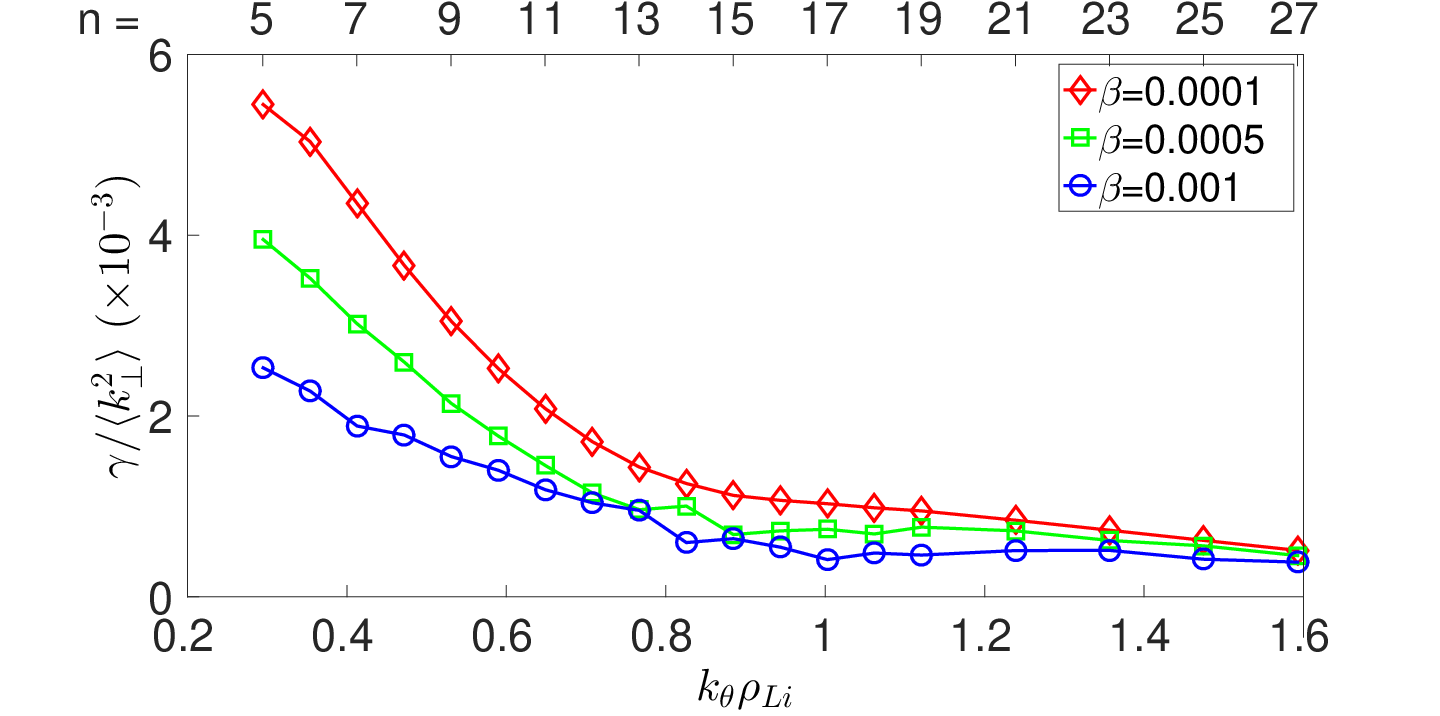}
\caption{\textcolor{black}{ Mixing length estimate for transport coefficient $D_{ML}=\gamma/\langle k_{\perp}^2\rangle$ \wrt $k_\theta \rho_{Li}$. The upper x-axis shows the corresponding toroidal mode
number $n$.} Three values of $\beta=0.0001, 0.0005$ and $0.001$ are considered.} 
\label{fig13}
\end{figure}

\noindent
It would be interesting to see how the \emf perturbation affects the overall flux.
Since this is a linear simulation, one can use a simple mixing length 
estimation of the transport coefficient. For the purpose, we calculate the ratio
$D_{ML}=\gamma/\langle k_{\perp}^2 \rangle$. \textcolor{black}{Here, $\gamma$ is the growth rate for a given $n$ and the mode square
average of the perpendicular wave-vector $\langle k_{\perp}^2 \rangle$ \textcolor{black}{= $\langle \kappa^2 \rangle + \langle k_\theta^2\rangle$}. Calculation of the mode square average of radial and poloidal wave-vectors are being done using the following expressions.
$$\langle \kappa^2 \rangle = \frac{\Sigma_{(k,m)} \big|\kappa \phi \big|^2 + \big|\kappa A_{||} \big|^2}{\Sigma_{(k,m)} \big| \phi \big|^2 + \big| A_{||} \big|^2}$$
$$\langle k_\theta^2\rangle=\frac{\int dr \Sigma_m \big|\frac{m}{r}\phi_{(k,m)} \big|^2 + \big|\frac{m}{r}A_{||(k,m)} \big|^2}{\int dr \Sigma_m \big|\phi_{(k,m)} \big|^2 + \big|A_{||(k,m)} \big|^2}$$}
In Fig. \ref{fig13} we calculate the quantity $D_{ML}$ for 
each value of toroidal mode number and plot the same \wrt the toroidal mode 
number. We consider three cases of $\beta$ values, $\beta=0.0001, 0.0005$ and $0.001$ 
corresponding to the results shown in Fig. \ref{fig3}. It is observed in Fig. \ref{fig13} that the mixing length estimation of heat flux peaks
at the longest wavelength despite the fact that the SWITG mode exhibits the strongest
growth rates around $n=21$. This implies that most of the contribution to the total
flux comes from modes close to the conventional ITG. This is qualitatively in conformity
with the results observed in Ref.~\cite{sw9}. It is to be noted that with increasing
value of $\beta$ the magnitude of $\gamma/{\langle k_{\perp}^2\rangle}$ decreases
implying a reduction in the heat flux. This is consistent with the stabilization
effect observed in Fig. \ref{fig3} and Fig. \ref{fig8} with respect to increasing $\beta$. Thus
we conclude that increasing $\beta$ does not only reduce growth rates of the mode
but \textcolor{black}{might also reduce the overall heat flux}.
\section{Summary}
In the present work, we have carried out a systematic study of the electromagnetic effect on the SWITG mode along with the conventional ITG mode using a global 
linear gyrokinetic model. Although the \emf effect on the ITG mode is hitherto 
well known, this is the first study that investigates the \emf effects on the SWITG 
mode using global gyrokinetic simulations. We have calculated the real frequency and growth rate 
for the chosen equilibrium for three different values of $\beta$.
We also have shown the mode structures for both the ITG and the SWITG mode 
in the presence of finite $\beta$. We have found out the growth rates and 
real frequencies \wrt $\beta$ and with increasing $\eta_i$ for different 
values of $\beta$. \textcolor{black}{Finally, we have presented the mixing length estimation of the 
transport}. The main findings are as follows.
\begin{enumerate}
\item{The SWITG mode is stabilized by the electromagnetic effect. The real 
frequency of the SWITG mode is weakly affected by the \emf effect.}
\item The SWITG mode-structure is more localized both radially and poloidally
compared to the ITG mode.
\item With increasing $\beta$ the real frequency of the SWITG mode decreases
slightly while the growth rate decreases substantially.
\item For the range of $\eta_i$ values studied, the growth rate of the SWITG mode instability decreases with an increase in $\beta$ value.
\textcolor{black}{
\item The ratio of electromagnetic to electrostatic potentials increases with increasing $\beta$.
\textcolor{black}{
\item The contribution of $\langle \kappa \rho_{Li} \rangle$ value is significant, compared to that of $\langle k_\theta \rho_{Li} \rangle$, for the calculation of $\langle k_\perp \rho_{Li} \rangle$. Also, with the increase in $\beta$ value, this contribution increases. }
\item The mixing length estimate for transport reveals that although the linear
growth rates of the SWITG mode are comparable to those of the long wavelength branch
the heat flux is maximum at the long wavelength region. This implies that most of the
contribution to the ion heat flux still comes from the conventional ITG mode that
occurs at $k_{\theta}\rho_i \leqslant 1.0$.
\item The mixing length estimate for transport shows that with increasing $\beta$
the heat flux decreases. This is in conformity with the linear stabilization of the
mode with increasing $\beta$.
}
\end{enumerate}
As is well known,  in the large aspect ratio limit, the trapped electron fraction  goes as $\sqrt{2\epsilon}$ where $\epsilon$ is the inverse aspect ratio $a/R_0.$
   Thus for typical aspect ratio considered here, effect of trapped electrons on finite $\beta$ SWITG may become important.
   It would therefore make sense to investigate this effect by considering the high $\beta$ version of the code EMGLOGYSTO,
   wherein dynamics due to ($\phi,~ A_{||},~ A_{\perp} $) perturbations may be addressed. This will be attempted in near future.
\section{Acknowledgement}
All the simulations reported here are performed using the Udbhav cluster at the Institute for Plasma Research(IPR). This paper is dedicated to late Professor Jan Vaclavik who taught one of the authors (RG) gyrokinetic theory.
\section{Data Availability}
The data that support the findings of this study are available from the corresponding author upon reasonable request.
\appendix
\section{}
%\numberwithin{equation}{section}
The set of equations discussed in Section II are closed by invoking the {\it quasi-neutrality condition} and the component of
Amp{\`e}re's law parallel to {\bf B}: 
\begin{eqnarray}
\sum_{j} {\tilde n}_{j}({\bf r};{\omega}) \simeq 0 ;\;\;\;\;\;\frac{1}{\mu _0} \nabla _\bot^{2} {\tilde A}_ \parallel = - \sum_{j}{\tilde j}_{\parallel j} \label{eq_closure1}
\end{eqnarray}
Equation (\ref{eq_closure1}) defines a generalized eigenvalue problem 
with eigenvalue $\omega$. % and eigenvector ${\tilde \varphi}$. 
This eigenvalue problem is 
conveniently solved in Fourier space. By Fourier decomposing the potential in Eq.(\ref{eq_closure1}) and then 
taking Fourier transform, we obtain a convolution matrix in Fourier space. 
If we assume a hydrogen-like plasma with ions, 
electrons, we have:   
%%%%%%%%%%%%%%%%%%%%%%%%%%%%%%%%%%%%%%%%matrix begins here%%%%%%%%%%%%%%%%%%%%%%%%%%%%%%%%%%%%%%%%%%%%%%5
%%%%------- Equation Environment -------%%%%
%\begin{widetext}
\begin{eqnarray}
\sum_{\bf k'} 
\left(
 \begin{array}{rr}
%\begin{matrix}
 \sum_{j=i,e}{\cal M}^{j}_{\tilde{\varphi} \tilde{\varphi},{\bf k},{\bf k'}} & \sum_{j=i,e}{\cal M}^{j}_{\tilde{\varphi} \tilde{A}_\parallel,{\bf k},{\bf k'}} \\
\sum_{j=i,e}{\cal M}^{j}_{\tilde{A}_\parallel \tilde{\varphi} ,{\bf k},{\bf k'}} & \sum_{j=i,e}{\cal M}^{j}_{\tilde{A}_\parallel \tilde{A}_\parallel,{\bf k},{\bf k'}}
\end{array}
\right)
%\end{matrix} 
%\begin{pmatrix}
\left(
 \begin{array}{r}
 {\tilde \varphi}_{\bf k'} \\
  {\tilde A}_{\parallel , \bf k'} \\
\end{array}
\right)
%\end{pmatrix} 
= 0
%%\]
%%$$
\end{eqnarray}
%\end{widetext}
%%%%------- Equation Environment -------%%%%
%%%%%%%%%%%%%%%%%%%%%%%%%%%%%%%%%%%%%%%%matrix ends here%%%%%%%%%%%%%%%%%%%%%%%%%%%%%%%%%%%%%%%%%%%%%%5
where ${\bf k} = (\kappa,m)$ and ${\bf k'}=(\kappa',m')$. Note that we have
2 species: passing ions ({\it i}) and passing electrons ({\it e}). In the following, we discuss in detail the formulation for
passing species. %For trapped ions, the reader is referred to \cite{sb_thesis}.
The Laplacian for the parallel 
component of Amp{\`e}re's law is also included in the appropriate matrix elements. The submatrices $ {\cal M} $ are symmetric about the diagonal. With the following definitions,
$\Delta\rho=\rho_{u}-\rho_{l}$ (upper and lower radial limits), $\Delta_{\kappa} = \kappa-\kappa'$ and $\Delta_{m} = m-m'$ matrix elements are :
%%%%%%%%%%%%%%%%%%%%%%1st row, ions%%%%%%%%%%%%%%%%%%%%%%%%%%%%%%%%%%%%%%%%%%%%%%%%%%%%%%%%%%%%%%%%%
\begin{eqnarray}
%&\mbox{}& 
&\nonumber 
{\cal M}^{i}_{\tilde{\varphi} \tilde{\varphi},{\bf k},{\bf k'}} &= 
%{\cal M}^{i}_{{\bf k},{\bf k'}} &=
\frac{1}{\Delta r }\int_{ r_{l}}^{ r_{u}} d{ r}\exp(-i \Delta_{\kappa} r) \times 
 \Bigg[ \alpha_{p} \delta_{mm'} + \exp(i \Delta_{m}{\bar\theta}) \sum_{p} {\hat I}^{0}_{p,i} \Bigg] , \\
%%%%%%%%%%%%%%%%%%%%%%%%%%%%%%%%%%%%%%%%%%%%%%%%%%%%%%%%
&\nonumber 
{\cal M}^{i}_{\tilde{\varphi} \tilde{A}_\parallel ,{\bf k},{\bf k'}}&= 
- \frac{1}{\Delta r}\int_{r_{l}}^{r_{u}} d{r}\exp(-i \Delta_{\kappa}r) \times 
 \Bigg[ \exp(i \Delta_{m}{\bar\theta}) \sum_{p} {\hat I}^{1}_{p,i} \Bigg] , \\
%%%%%%%%%%%%%%%%%%%%%%%%%%%%%%%%%%%%%%%%%%%%%%%%%%%%%%%%
&\nonumber 
{\cal M}^{e}_{\tilde{\varphi} \tilde{\varphi},{\bf k},{\bf k'}} &= 
%{\cal M}^{i}_{{\bf k},{\bf k'}}&=
\frac{1}{\Delta r}\int_{r_{l}}^{r_{u}} d{r}\exp(-i \Delta_{\kappa}r) \times 
 \Bigg[ \frac{\alpha_{p} \delta_{mm'}}{\tau(r)} + \frac{\exp(i \Delta_{m}{\bar\theta})}{\tau(r)} \sum_{p} {\hat I}^{0}_{p,e} \Bigg] , \\
%%%%%%%%%%%%%%%%%%%%%%%%%%%%%%%%%%%%%%%%%%%%%%%%%%%%%%%%
\nonumber 
&{\cal M}^{e}_{\tilde{\varphi} \tilde{A}_\parallel ,{\bf k},{\bf k'}} &= 
- \frac{1}{\Delta r}\int_{r_{l}}^{r_{u}} d{r}\exp(-i \Delta_{\kappa}r) \times 
 \Bigg[ \frac{ \exp(i \Delta_{m}{\bar\theta})}{\tau(r)} \sum_{p} {\hat I}^{1}_{p,e} \Bigg] , \\
%%%%%%%%%%%%%%%%%%%%%%%%%%%%%%%%%%%%%%%%%%%%%%%%%%%%%%%%
%%%%%%%%%%%%%%%%%%%%%%%%%%%%%%%%%%%%%%%%%%%%%%%%%%%%%%%%
&\nonumber 
{\cal M}^{i}_{\tilde{A}_\parallel \tilde{A}_\parallel,{\bf k},{\bf k'}} &=
\frac{1}{\Delta r}\int_{r_{l}}^{r_{u}} d{r}\exp(-i \Delta_{\kappa}r) \times 
 \Bigg[ \exp(i \Delta_{m}{\bar\theta}) \sum_{p} {\hat I}^{2}_{p,i} \Bigg] \\
%%%%%%%%%%%%%%%%%%%%%%%%%%%%%%%%%%%%%%%%%%%%%%%%%%%%%%%%
& &\nonumber 
- \frac{1}{\Delta r}\int_{r_{l}}^{r_{u}} d{r}\exp(-i \Delta_{\kappa}r) \times 
% \Bigg[ 
 \left( \kappa^{\prime 2} + \frac{m^{ \prime 2}}{r ^{2}} \right) \left( \frac{T_i(r)}{q_i^2 N \mu_0} \right)  , \\
%%%%%%%%%%%%%%%%%%%%%%%%%%%%%%%%%%%%%%%%%%%%%%%%%%%%%%%%
%\nonumber 
&{\cal M}^{e}_{\tilde{A}_\parallel \tilde{A}_\parallel,{\bf k},{\bf k'}} &=
\frac{1}{\Delta r}\int_{r_{l}}^{r_{u}} d{r} \frac{\exp(-i \Delta_{\kappa}r)}{\tau(r)} \times 
 \Bigg[ \exp(i \Delta_{m}{\bar\theta}) \sum_{p} {\hat I}^{2}_{p,e} \Bigg]  \label{eq_Matrix}
%%%%%%%%%%%%%%%%%%%%%%1st row, electrons%%%%%%%%%%%%%%%%%%%%%%%%%%%%%%%%%%%%%%%%%%%%%%%%%%%%%%%%%%%%%%%%%
\end{eqnarray}
%%%%%%%%%%%%%%%%%%%%%%%%%%%%%%%%%%%%%%%%%%%%%%%%%%%%%%%%
where
%%%%%%%%%%%%%%%%%%%%%%%%%%%%%%%%%%%%%%%%%%%%%%%%%%%%%%%%
\begin{eqnarray}
%%%%%%%%%%%%%%%%%%%%%%%%%%%%%%%%%%%%%%%%%%%%%end of matrix elements%%%%%%%%%%%%
&\mbox{}&\nonumber {\hat I}^{l}_{p,j}=\frac{1}{\sqrt{2\pi} v^{3}_{th,j}(r)}\int_{-v_{max,j}(r)}^{v_{max,j}(r)}v^{l}_{||} dv_{||} \exp{\left(-\frac{v^{2}_{||}}{v^{2}_{th,j}(r)}\right)}
\left\{\frac{N^{j}_{1}{I}^{\sigma}_{0,j}-N^{j}_{2}{I}^{\sigma}_{1,j}}{D_{1}^{\sigma,j}}\right\}_{p'=p-(m-m')} ,
\end{eqnarray}
%%%%%%%%%%%%%%%%%%%%%%%%%%%%%%%%%%%%%%%%%%%%%%%%%%%%%%%%%%%%%%
%%%%%%%%%%%%%%%%%%%%%%%%%%%%%%%%%%%%%%%%%%%%%%%%%%%%%%%%%%%%%%
\begin{eqnarray}
\nonumber{I}^{\sigma}_{n,j}=\int_{0}^{v_{\perp max,j}(r)}v^{2n+1}_{\perp} dv_{\perp} \exp{\left(-\frac{v^{2}_{\perp}}{2v^{2}_{th,j}(r)}\right)} J^{2}_{0}(x_{Lj}) J_{p}(x^{'\sigma}_{tj}) J_{p'}(x^{'\sigma}_{tj}) ~, \label{eq_Icap}
\end{eqnarray}
We have introduced the following definitions: $\epsilon$ is the inverse aspect ratio,
$v_{max,j}(r) $ is the upper cutoff speed (considered in the numerical implementation) of the species $j$, $v_{\perp max,j(\rho)}={\it min}(v_{||}/\sqrt{\epsilon},v_{max,j})$ which is ``trapped particle exclusion'' from 
$\omega$ independent perpendicular velocity integral
${I}^{\sigma}_{n,j}$; $\alpha_{p}=1 - \sqrt{\epsilon/(1+\epsilon)}$ is the 
fraction of passing particles; ${\hat I}^{l}_{p,j}$, 
is $\omega-{\rm dependent}$ parallel integrals; 
$x^{\sigma}_{tj}=k_{\perp}\xi_{\sigma}$,
$N^{j}_{1}=\omega - w_{n,j}\left[1 + (\eta_{j}/2)(v^{2}_{||}/v^{2}_{th,j})-3)\right]$;  $N^{j}_{2}= w_{n,j}\eta_{j}/(2 v^{2}_{th,j})$
and $D^{\sigma,j}_{1}=<w_{t,j}(\rho)>(n q_{s} - m'(1-p)(\sigma v_{||}/v_{th,j}) - \omega$ where $<w_{t,j}(\rho)> =v_{th,j}(\rho)/(r q_s)$ is the average {\it transit frequency} of the species $j$. 

\bibliographystyle{apsrev4-1}
%\section*{References}
 \bibliography{Revised_Manuscript_RGanesh}

\end{document}